\documentclass[%
reprint,
preprintnumbers,
superscriptaddress,
 amsmath,amssymb,
 aps,
prx]{revtex4-2}

\usepackage{graphicx}
\usepackage{dcolumn}
\usepackage{bm}
\usepackage{amsmath}
\usepackage{comment}
\usepackage[hidelinks]{hyperref}
\usepackage[normalem]{ulem}
\usepackage{braket}

\usepackage[dvipsnames,usenames]{xcolor}

\begin{document}

\preprint{FERMILAB-PUB-26-0007-ETD-PPD}

\title{A First Demonstration of the SQUAT Detector Architecture: \\ Direct Measurement of Resonator-Free Charge-Sensitive Transmons}

\author{H.~Magoon}
\email[]{hwmagoon@stanford.edu}
\affiliation{SLAC National Accelerator Laboratory, Menlo Park, California 94025, USA}
\affiliation{Kavli Institute for Particle Astrophysics and Cosmology, Stanford University, Stanford, CA 94035, USA}
\affiliation{Department of Physics, Stanford University, Stanford, CA 94035, USA}

\author{T.~Aralis}
\email[]{taralis@slac.stanford.edu}
\affiliation{SLAC National Accelerator Laboratory, Menlo Park, California 94025, USA}
\affiliation{Kavli Institute for Particle Astrophysics and Cosmology, Stanford University, Stanford, CA 94035, USA}

\author{T. Dyson}
\affiliation{SLAC National Accelerator Laboratory, Menlo Park, California 94025, USA}
\affiliation{Kavli Institute for Particle Astrophysics and Cosmology, Stanford University, Stanford, CA 94035, USA}
\affiliation{Department of Physics, Stanford University, Stanford, CA 94035, USA}

\author{J.~Anczarski}
\affiliation{SLAC National Accelerator Laboratory, Menlo Park, California 94025, USA}
\affiliation{Kavli Institute for Particle Astrophysics and Cosmology, Stanford University, Stanford, CA 94035, USA}
\affiliation{Department of Physics, Stanford University, Stanford, CA 94035, USA}

\author{D. Baxter}
\affiliation{Fermi National Accelerator Laboratory, Batavia, IL 60510, USA}
\affiliation{Department of Physics \(\&\) Astronomy, Northwestern University, Evanston, IL 60208, USA}

\author{G. Bratrud}
\affiliation{Department of Physics \(\&\) Astronomy, Northwestern University, Evanston, IL 60208, USA}
\affiliation{Fermi National Accelerator Laboratory, Batavia, IL 60510, USA}

\author{R. Carpenter}
\affiliation{SLAC National Accelerator Laboratory, Menlo Park, California 94025, USA}
\affiliation{Kavli Institute for Particle Astrophysics and Cosmology, Stanford University, Stanford, CA 94035, USA}
\affiliation{Department of Applied Physics, Stanford University, Stanford, CA 94035, USA}

\author{S. Condon}
\affiliation{SLAC National Accelerator Laboratory, Menlo Park, California 94025, USA}
\affiliation{Kavli Institute for Particle Astrophysics and Cosmology, Stanford University, Stanford, CA 94035, USA}
\affiliation{Department of Physics, Stanford University, Stanford, CA 94035, USA}

\author{A. Droster}
\affiliation{SLAC National Accelerator Laboratory, Menlo Park, California 94025, USA}
\affiliation{Kavli Institute for Particle Astrophysics and Cosmology, Stanford University, Stanford, CA 94035, USA}

\author{E. Figueroa-Feliciano}
\affiliation{Department of Physics \(\&\) Astronomy, Northwestern University, Evanston, IL 60208, USA}
\affiliation{Fermi National Accelerator Laboratory, Batavia, IL 60510, USA}

\author{C.W.~Fink}
\affiliation{Institute for Quantum \& Information Sciences, Syracuse University, Syracuse, NY 13244, USA}
\affiliation{Department of Physics, Syracuse University, Syracuse, NY 13244, USA}

\author{S.~Harvey}
\affiliation{SLAC National Accelerator Laboratory, Menlo Park, California 94025, USA}

\author{A.~Simchony}
\affiliation{SLAC National Accelerator Laboratory, Menlo Park, California 94025, USA}
\affiliation{Kavli Institute for Particle Astrophysics and Cosmology, Stanford University, Stanford, CA 94035, USA}
\affiliation{Department of Physics, Stanford University, Stanford, CA 94035, USA}

\author{Z.J.~Smith}
\affiliation{SLAC National Accelerator Laboratory, Menlo Park, California 94025, USA}
\affiliation{Kavli Institute for Particle Astrophysics and Cosmology, Stanford University, Stanford, CA 94035, USA}
\affiliation{Department of Applied Physics, Stanford University, Stanford, CA 94035, USA}

\author{S. Stevens}
\affiliation{SLAC National Accelerator Laboratory, Menlo Park, California 94025, USA}
\affiliation{Kavli Institute for Particle Astrophysics and Cosmology, Stanford University, Stanford, CA 94035, USA}

\author{N. Tabassum}
\affiliation{SLAC National Accelerator Laboratory, Menlo Park, California 94025, USA}
\affiliation{Kavli Institute for Particle Astrophysics and Cosmology, Stanford University, Stanford, CA 94035, USA}

\author{B.A. Young}
\affiliation{Santa Clara University, Santa Clara, CA 95053, USA}

\author{C.P.~Salemi}
\affiliation{SLAC National Accelerator Laboratory, Menlo Park, California 94025, USA}
\affiliation{Kavli Institute for Particle Astrophysics and Cosmology, Stanford University, Stanford, CA 94035, USA}
\affiliation{University of California Berkeley, Berkeley, CA 94720, USA}
\affiliation{Lawrence Berkeley National Laboratory, Berkeley, CA 94720, USA}

\author{K. Stifter}
\affiliation{SLAC National Accelerator Laboratory, Menlo Park, California 94025, USA}
\affiliation{Kavli Institute for Particle Astrophysics and Cosmology, Stanford University, Stanford, CA 94035, USA}

\author{D.I. Schuster}
\affiliation{Department of Applied Physics, Stanford University, Stanford, CA 94035, USA}

\author{N.A. Kurinsky}
\affiliation{SLAC National Accelerator Laboratory, Menlo Park, California 94025, USA}
\affiliation{Kavli Institute for Particle Astrophysics and Cosmology, Stanford University, Stanford, CA 94035, USA}



\date{\today}

\begin{abstract}
The Superconducting Quasiparticle-Amplifying Transmon~(SQUAT) is a new sensor architecture for THz~(meV) detection based on a weakly charge-sensitive transmon directly coupled to a transmission line.  In such devices, energy depositions break Cooper pairs in the qubit capacitor islands, generating quasiparticles. Quasiparticles that tunnel across the Josephson junction change the transmon qubit parity, generating a measurable signal.  In this paper, we present the design of first-generation SQUATs and demonstrate an architecture validation.  We summarize initial characterization measurements made with prototype devices, comment on background sources that influence the observed parity-switching rate, and present experimental results showing simultaneous detection of charge and quasiparticle signals using aluminum-based SQUATs.
\end{abstract}

\maketitle

\section{Introduction}\label{sec:intro}

Within the rapidly developing field of quantum computing, superconducting circuits remain a leading platform for qubit realization. Early superconducting qubits were highly susceptible to decoherence via `quasiparticle poisoning', the process through which energy depositions break Cooper pairs in superconducting metals to produce Bogoliubov quasiparticles. This increase in quasiparticle density promotes tunneling across the qubit Josephson junction and causes both state decay and dephasing (see Ref.~\cite{siddiqiReview} and references therein).

The energy required to break a single Cooper pair (producing two quasiparticles) is material dependent and equivalent to twice the superconducting band gap~($\Delta$). Aluminum (Al), a common choice for superconducting circuits, has $\Delta \approx200\,\mu$eV, making it susceptible to excess quasiparticle generation from sub-meV sources. Modern qubit designs use gap engineering to mitigate this sensitivity, reducing tunneling and dramatically improving coherence~\cite{McEwen_2021}. Qubits with minimal gap engineering remain interesting for sensing applications, where generated quasiparticles indicate the presence of signal and the small scale of $\Delta$ provides a low energy-detection threshold. There are a variety of uses for meV/THz sensing, including dark-matter direct detection~\cite{chou2023}, nuclear monitoring~\cite{Bowen_2020,Akindele_2021}, and photon down-conversion in quantum networks~\cite{ahmed2025thz}.

Several existing `pair-breaking' sensor architectures are currently in use or being developed for rare event searches. These architectures include the transition edge sensor~(TES)~\cite{cabrera2000,irwin_hilton,Ren_2021,Anthony_Petersen_2025}, the microwave kinetic inductance detector~(MKID)~\cite{jonas_arcmp,primaKID}, the superconducting nanowire single-photon detector~(SNSPD)~\cite{Marsili_2011,Luskin_2023}, the quantum capacitance detector~(QCD)~\cite{echternach_2018}, and the related quantum parity detector~(QPD)~\cite{ramanathan2024}. These devices have varying levels of maturity and include tradeoffs in energy threshold, energy resolution, noise performance, and potential for multiplexing. None of these existing architectures have yet demonstrated the ability to resolve single THz photons or meV phonons, although the QCD has achieved single THz photon counting at a fixed frequency~\cite{echternach_2018}.

\begin{figure*}[!t]
    \centering
    \begin{minipage}[c]{0.40\linewidth}
        \centering
        \includegraphics[width=\linewidth]{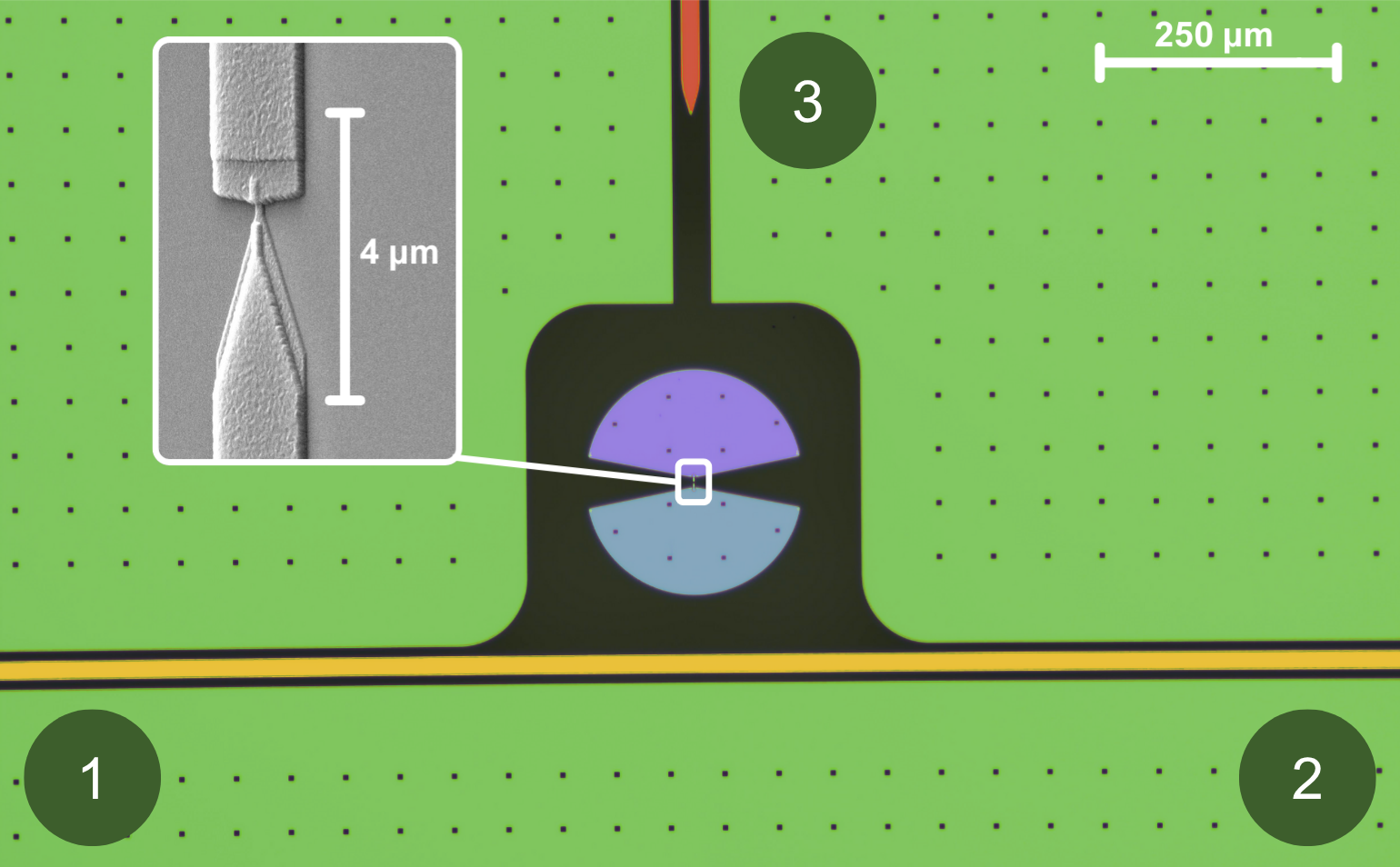}
    \end{minipage}%
    \begin{minipage}[c]{0.50\linewidth}
        \centering
        \includegraphics[width=\linewidth]{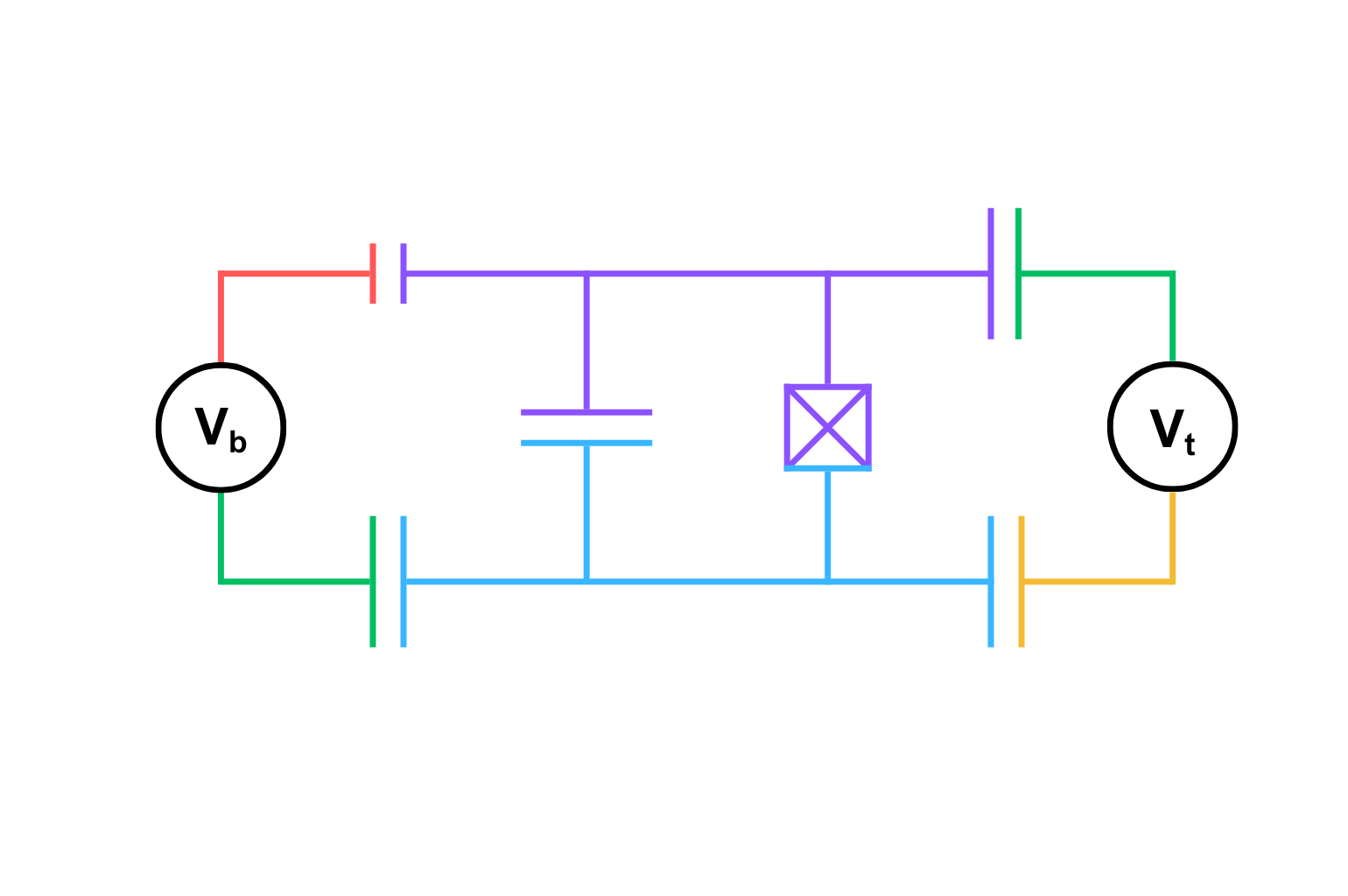}
    \end{minipage}
    \begin{minipage}[t]{0.90\linewidth}
        \centering
         \includegraphics[width=\linewidth]{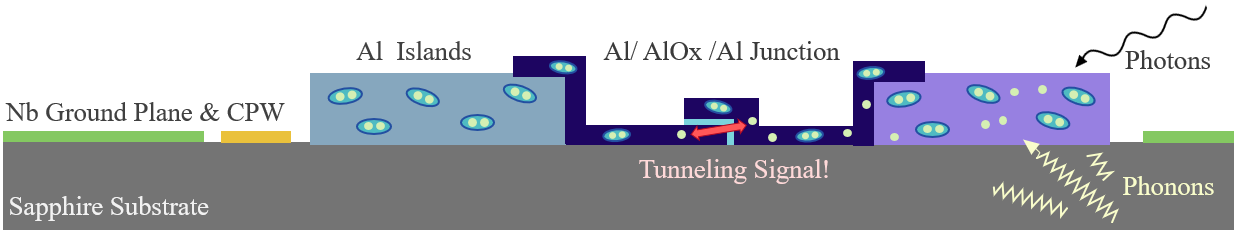}
    \end{minipage}
    \caption{\textbf{Top Left}: 
    Micrograph of one of three SQUATs on a prototype chip. The SQUAT is centered in the image, with fin-shaped islands (purple and blue) surrounded by a ground plane (green).  The SQUAT is strongly coupled to the transmission line (yellow) and weakly coupled to the charge bias line (red). The transmission line can be used to measure transmission past the SQUAT ($S_{21}$), while the charge line can be used to measure transmission through the SQUAT ($S_{23}$), making this a 3-port equivalent network.  An SEM of the junction is shown in the inset image.
    \textbf{Top Right}: 
    The circuit diagram for a single SQUAT. The SQUAT characteristics are determined by the capacitance between the islands, to ground, and to the feedline, as well as the Josephson energy ($E_J$) of the junction. The SQUAT can be DC biased through either the transmission line or the charge line, with voltages $V_t$ and $V_b$ respectively.
    \textbf{Bottom}: 
    Diagrammatic representation of the SQUAT mechanism for detecting photon and phonon events. Phonons from the crystalline substrate or directly absorbed photons can break Cooper pairs in either island, increasing the quasiparticle density. Quasiparticles that drift near the junction have some probability to tunnel, producing measurable signal.}\label{fig:designOverview}
\end{figure*}

In Ref.~\cite{fink2024}, members of this team presented the concept for a novel qubit-based pair-breaking sensor, the superconducting quasiparticle-amplifying transmon~(SQUAT). SQUATs employ a weakly charge-sensitive transmon directly coupled to a transmission line. In the weakly charge-sensitive regime, the first excited-state energy varies periodically with offset charge~\cite{koch}. A half period is equivalent to a single elementary charge, meaning a quasiparticle that tunnels across the Josephson junction can produce a discrete jump in excitation energy. Such a change is detectable by directly probing the qubit via the transmission line.  Repeated quasiparticle tunneling produces two observable `parity states' differentiated by whether an even or odd number of quasiparticles have crossed the junction at the time of measurement. Energy depositions that increase the quasiparticle density in a SQUAT can therefore produce a detectable increase in the rate of parity-switching. Such events can be directly probed without the use of a readout resonator, allowing for closer pixel packing in sensor arrays and avoiding the loss of detectable phonon energy to insensitive resonator metal.  An overview of the SQUAT geometry is shown in Fig.~\ref{fig:designOverview}.

In conventional transmon designs, gap engineering is used to preferentially trap quasiparticles on the low-gap side of the junction, reducing the probability of tunneling. In SQUAT designs, gap engineering can instead be used to enhance sensitivity by trapping quasiparticles near the junction, increasing the tunneling rate for a given energy deposition. Trapping is effected by fabricating the junction with lower-gap material than the capacitor islands, so quasiparticles downconverted near the junction no longer have sufficient energy to return to the islands. Quasiparticles trapped near the junction have a higher likelihood to tunnel (possibly multiple times) before recombination, thus increasing the number of tunneling signals per event-generated quasiparticle. The downconversion process also amplifies quasiparticle number, increasing the probability that broken pairs in the higher-gap absorber produce measurable tunneling in the junction.

In this paper, we present results from the first prototype devices, made using Al/AlOx/Al junctions on sapphire to demonstrate SQUAT readout and design fidelity. Note that these initial devices were made with entirely Al islands and thus involve no island-to-junction trapping. In Sec.~\ref{sec:design}, we detail SQUAT design parameters and discuss the values chosen for the prototype detectors covered in this paper. In Sec.~\ref{sec:deviceOnly}, we describe measurements of the steady-state properties of the SQUATs, such as transmission characteristics and coherence times. In Sec.~\ref{sec:parity_meas}, we demonstrate the use of SQUATs as sensors, including measurements of the dependence of parity-switching rate on conditions such as temperature and loading from thermal radiation. Lastly, in Sec.~\ref{sec:discussion}, we discuss future directions and near-term applications for both phonon- and photon-coupled SQUAT sensors.

\section{SQUAT Design} \label{sec:design}

The SQUAT design is based on the standard transmon qubit architecture~\cite{koch}, modified to maximize external quantum efficiency for photon and phonon collection. In a SQUAT, the qubit is capacitively coupled to the transmission line directly, rather than via an intermediate readout resonator. This enables continuous high-bandwidth single-tone measurement of dispersion and improves detection efficiency by reducing the chip surface area occupied by non-sensor phonon-absorbing material. Additionally, the SQUAT's capacitor islands are designed to enhance quasiparticle diffusion and direct photon coupling. Implementing these changes requires careful optimization to balance qubit energy scales, coupling efficiency, and quasiparticle collection.

The SQUAT transmon intentionally retains weak charge sensitivity, characterized by the ratio of the Josephson energy ($E_J$) to the capacitive charging energy ($E_C$).  With $E_J/E_C \sim 25$, the device operates beyond the Coulomb blockade regime while maintaining a measurable maximum charge dispersion ($2\chi_0$) between even and odd parity states, enabling detection of quasiparticle-tunneling events.

The dispersion and SQUAT transition frequency ($f_0$) are related to $E_J$ and $E_C$ via the standard transmon relations: 
\begin{align}\label{eqs:squatParams}
    hf_0 &\approx \sqrt{8E_CE_J} - E_C,  \\ 
    \frac{2\chi_0}{f_0}&\approx e^{-\sqrt{8\frac{E_J}{E_C}}}\left[60.7\left(\frac{E_J}{E_C}\right)^{3/4}+5.37\left(\frac{E_J}{E_C}\right)^{1/4}\right] \nonumber
\end{align}

\subsection{Design Considerations}

Following the coherent-response model of Ref.~\cite{sultanov} and detailed in App.~\ref{app:hamiltonian}, we find that, for a given readout power, the highest fidelity parity readout occurs when the parity-state dispersion satisfies
\begin{equation}\label{eq:dispersion}
    2\chi = 2\gamma\sqrt{1+\frac{2\Gamma_n}{\gamma}} = \frac{2\pi f_0}{Q_r}\sqrt{1+4\frac{\Gamma_nQ_r}{2\pi f_0}}
\end{equation}
where $\chi$ is the charge dispersion at the time of readout, $Q_r$ is the total quality factor of the resonance, $\Gamma_n$ is the `photon interrogation rate' (the average rate of photons arriving at the SQUAT), and $\gamma = 2\pi f_0 / (2Q_r)$ is the total qubit decoherence rate (the inverse of the qubit's total dephasing time $T_2$). To describe the SQUAT’s dynamics, we express $\gamma$ in terms of its individual components:
\begin{equation}
    \gamma \equiv \frac{1}{2}\Gamma_r + \Gamma_\phi
\end{equation}
The physical decay rates contributing to decoherence are defined as follows:
\begin{itemize}
    \item $\Gamma_r$ is the total radiative decay rate, which can be further broken down by loss channel:
    \begin{equation}
        \Gamma_r \equiv \Gamma_c + \Gamma_l
    \end{equation}
    \item $\Gamma_c$ is the radiative decay to the feedline that provides the measured signal
    \item $\Gamma_l$ describes the radiative loss rate to everything but the feedline
    \item{$\Gamma_\phi$} is the pure dephasing rate, accounting for phase noise processes
\end{itemize}
In the low-power limit (where $2\Gamma_n/\gamma \ll 1$), Eq.~\ref{eq:dispersion} reproduces the optimal fidelity from Ref.~\cite{fink2024}. The two results deviate at higher powers, where Eq.~\ref{eq:dispersion} appropriately treats the qubit as a two-level-system rather than using a semi-classical model. 

As derived in App.~\ref{app:fidelity}, the final state variance for optimal phase readout can be expressed as
\begin{equation}\label{eq:stateVariance_main}
    \sigma_s^2 = 16\beta \left[\frac{2\gamma}{\Gamma_c}\right]^2 \frac{2}{\gamma}f_{bw}\eta(\beta^{-1}) \geq 64Q_r\frac{k_b T_n}{2\pi hf_0^2}f_{bw} \eta(\beta)
\end{equation}
where $\beta = k_B T_n / h f_0$ is the mean photon occupancy for SQUAT noise temperature $T_n$, and $f_{bw}$ is the readout bandwidth.  The inequality highlights that, for fixed bandwidth, the minimum achievable state variance decreases with decreasing $Q_r$, motivating the use of strongly coupled (low-$Q$) SQUATs for fast, high-fidelity readout.  However, reducing $Q_r$ also limits potential multiplexing density, leading to a tradeoff between bandwidth and channel count.  Targeting $Q_r \sim 10^3$ and $2\chi_0 \sim 10$ MHz allows multiplexing factors of order $10^2$ for GHz-scale $f_0$.

For our Al/AlOx/Al junctions on sapphire, we achieved $E_J/h \approx 16$ GHz.  The charging energy $E_C/h \sim 600-700$~MHz was tuned through the island geometry, with total capacitance verified by finite-element simulation (see App.~\ref{app:design}).

\subsection{Prototype Design}

We used a three-SQUAT design to demonstrate multiplexability with limited risk of resonance overlap. A colorized schematic of one SQUAT is shown in Fig.~\ref{fig:designOverview}, and the full chip and package layout are discussed in App.~\ref{app:device_env}.  These devices were designed for readout bandwidth up to $f_{bw}\sim {f_0}/{Q_r}\sim 1\,\text{MHz}$. Individual SQUATs were designed to have dispersions $2\chi_0 \sim 10\,\text{MHz}$, and a SQUAT-to-SQUAT frequency separation of approximately 25\,MHz to avoid dispersive resonance crossings and to minimize crosstalk.  The complete set of expected SQUAT properties can be found in App.~\ref{app:design}.

With the goal of architecture validation, the first prototype devices were produced with no quasiparticle trapping.  Instead, aluminum was used for both the islands and junctions.  A niobium (Nb) ground plane covers the majority of the chip, with flux holes spaced 55\,$\mu$m apart to reduce spurious vortex loss.  A Nb coplanar waveguide transmission line is coupled to all three SQUATs, and each SQUAT has a dedicated weakly coupled Nb charge-bias line. 

The Al junction leads were deposited in two thicknesses, 45 and 115\,nm, resulting in a small superconducting gap asymmetry ($\delta\Delta$) across the junction. These weakly gap-engineered junctions, designed in the $\delta \Delta < h f_0$ regime, exhibit a smaller gap difference than designs optimized to suppress quasiparticle poisoning~\cite{McEwen_2021}. We chose this design to explore the tradeoff between quasiparticle tunneling efficiency and background parity-switching rate; the observed implications of this choice are discussed in Sec.~\ref{sec:parity_meas}.

\section{Steady-State Properties}\label{sec:deviceOnly}

To validate the SQUAT architecture, we first characterized each device's steady-state properties, including continuous-wave (CW) resonance parameters and qubit coherence times. Such measurements are required to optimize SQUAT readout and fidelity, independent of event- or background-induced tunneling. The experimental setup is detailed in App.~\ref{app:exp_setup}.

\subsection{Frequency-Dependent Transmission Measurements}

When a CW tone (with duration longer than $T_2$) is applied, the SQUAT is driven to a mixed state. As detailed in App.~\ref{app:hamiltonian}, the response can be represented as a two-level system coupled to a bath of photons representing the transmission line \cite{Peropadre_2013, emely_2021}.  In this model, the frequency-dependent transmission of each parity band can be fit using Eq.~\ref{eq:S21Squat} to extract decay rates and on-chip tone power:
\begin{equation}\label{eq:S21Squat}
    S_{21}(f) = 1 - \frac{\Gamma_c}{2\gamma} \frac{1-i\frac{2\pi\delta f}{\gamma}}{1 + \left(\frac{2\pi\delta f}{\gamma} \right)^2 + \frac{2\Gamma_n}{\gamma}}
\end{equation}
where $\delta f = f - f_r$ is the frequency detuning, $f_r=f_0\pm\chi$ is the parity-dependent resonance frequency, $\Gamma_n = 2P_r/hf_r$ is photon interrogation rate, and $P_r$ is readout power at the device.

At low drive power ($2\Gamma_n/\gamma \ll 1$), the response becomes Lorentzian (the approximation used in \cite{fink2024}). This limit roughly corresponds to one or fewer incident readout photons per SQUAT photon decay. We use this limit to extract relationships between decay rates and quality factors (see Eq.~\ref{Quality_factors} in App.~\ref{app:hamiltonian}). We can then rewrite the low-power limit:
\begin{equation}\label{eq:full_response}
   P_r \ll \frac{\pi h f_r^2}{4Q_r}.
\end{equation}

\begin{figure*}
    \centering
    \includegraphics[width=0.9\linewidth]{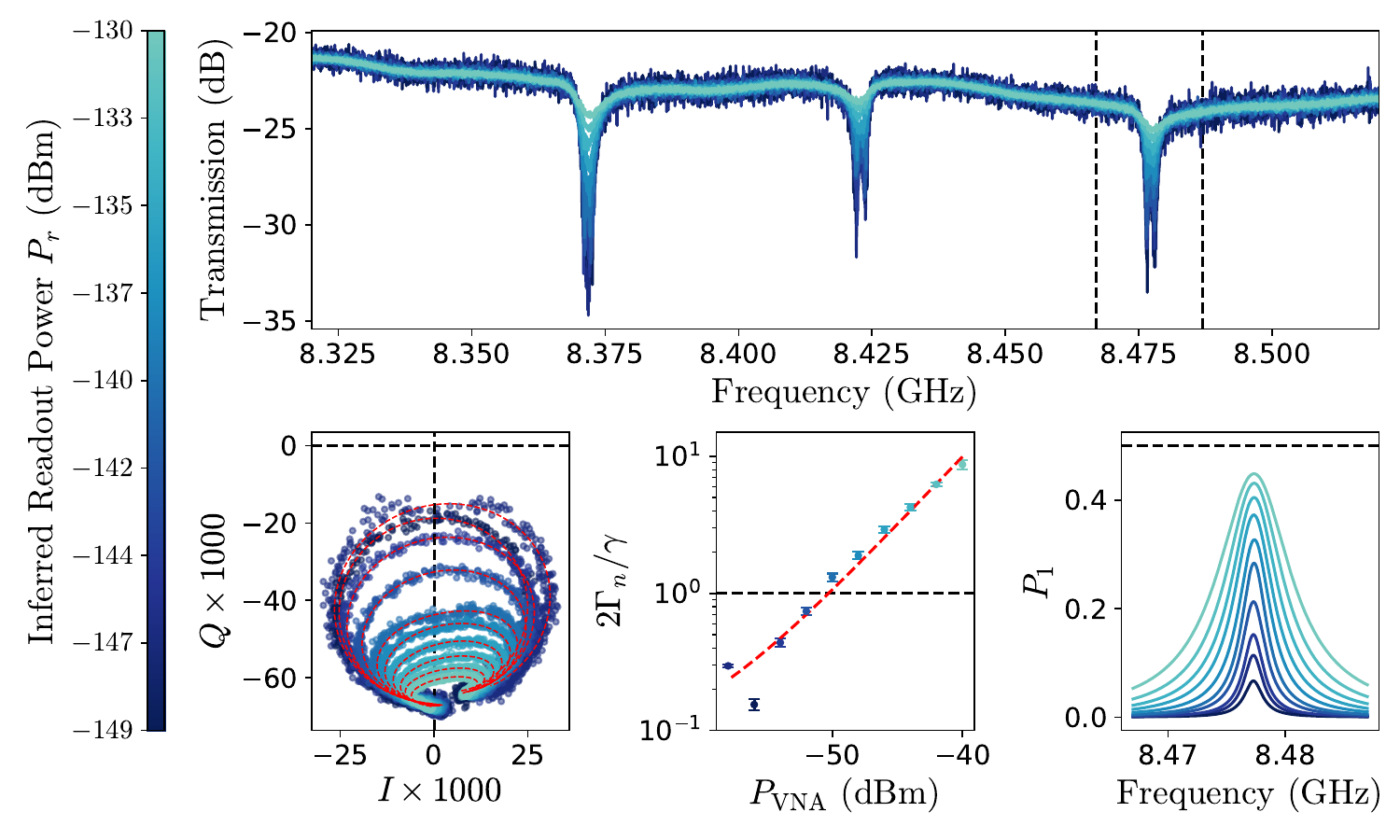}
    \caption{\textbf{Top:} The magnitude of transmission ($S_{21}$) for three SQUATs on a single chip at multiple readout powers. \textbf{Bottom~Left:} Complex $S_{21}$ for one of the above SQUATs with fits at each power. The fit interrogation rate~($\Gamma_n$) is used to calculate the inferred on-chip readout powers~($P_r$) for the color axis. Compression of the resonance loop due to increased qubit excitation probability~($P_1$) is visible for higher powers. \textbf{Bottom~Center:} Fit $\Gamma_n$ normalized to the total qubit decoherence rate~($\gamma$) as a function of readout power from the VNA. \textbf{Bottom~Right:} $P_1$ as a function of readout frequency and power, calculated from Eq.~\ref{eq:P1} with fit results from Eq.~\ref{eq:S21Squat}.}
    \label{fig:fscan_all}
\end{figure*}

For the measured devices, we reach the low power regime at approximately:
\begin{align}\label{eq:Pr}
    P_{r} &\ll 3.3\cdot 10^{-17}\,\mathrm{W} \left[\frac{f_0}{8\;\mathrm{GHz}}\right]^2\left[\frac{1000}{Q_r}\right] \\
    &\approx -135\;\mathrm{dBm} +20\log_{10}\left[\frac{f_0}{8\;\mathrm{GHz}}\right]-10\log_{10}\left[\frac{Q_r}{1000}\right].
\end{align}

Fig.~\ref{fig:fscan_all} shows a representative VNA frequency scan for a three-SQUAT chip, taken over a range of readout powers.  As power increases above the single-photon limit, the resonance becomes shallow and less circular in the complex plane, consistent with the expected two-level response (see App.~\ref{app:hamiltonian}).  Fitting the data to Eq.~\ref{eq:S21Squat} yields $\Gamma_c$, $\gamma$, $f_r$, and $\Gamma_n$ as a function of VNA power ($P_\mathrm{VNA}$).  Subsequently fitting the slope of $\Gamma_n$ versus $P_\mathrm{VNA}$ provides an absolute calibration of on-chip readout power, analogous to the Stark-shift calibration used for conventional transmons \cite{Schuster_2005}.  A summary of fit parameters obtained without any fixed priors for all five measured qubits is given in Tab.~\ref{tab:fitValues}.

Results from fitting Eq.~\ref{eq:S21Squat} can also be applied to calculate the SQUAT's probability to be in the excited state ($P_1$) as a function of readout power and frequency (see App.~\ref{app:hamiltonian}).
\begin{equation}\label{eq:P1}
    P_1 = \frac{1}{2} \frac{\frac{2\Gamma_n}{\gamma}}{1 + \left(\frac{2\pi\delta f}{\gamma} \right)^2 + \frac{2\Gamma_n}{\gamma}} 
\end{equation}
Calculated $P_1(f)$ for each readout power are also shown in Fig.~\ref{fig:fscan_all}. Excitation is strongly peaked around the resonance and broadens for higher power.

\subsection{Charge Dependence}

The SQUAT exhibits a characteristic even-odd parity dispersion ($2\chi$), which varies periodically as a function of offset charge $n_g$ \cite{koch}:
\begin{equation}
    2\chi = 2\chi_0 \; \text{cos}(\pi n_g)
\end{equation}
where $2\chi_0$ is the maximum even–odd parity splitting set primarily by the energy ratio $E_J/E_C$.  When the measurement integration time exceeds the parity-switching time, both parity bands appear simultaneously in VNA-style transmission measurements (Fig.~\ref{fig:fscan_all}).

\begin{figure}[th!]
    \centering
    \includegraphics[width=1.0\columnwidth]{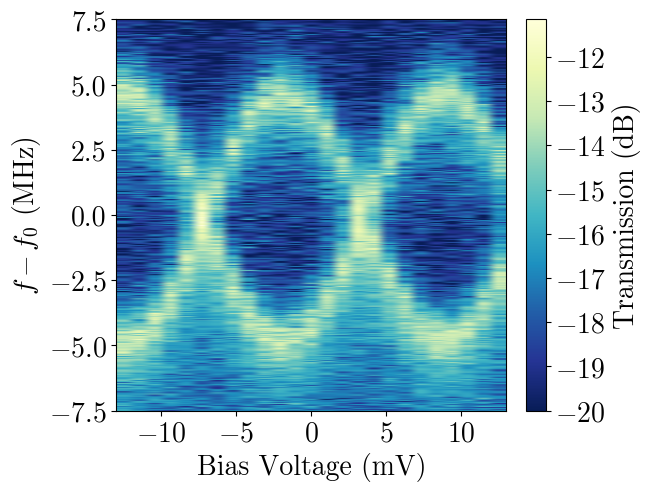}
    \caption{The magnitude of transmission ($S_{23}$) as a function of island charge. A DC bias voltage is applied on the transmission line, and a frequency scan is taken in the vicinity of the qubit.  The SQUAT dispersion~($2\chi$) varies periodically as a function of induced charge.  The voltages shown here are applied at the input to the fridge, along wiring shown in Fig.~\ref{fig:readout_chain}.}
    \label{fig:charge_scan}
\end{figure}

By applying a DC bias to the charge or transmission line ($V_b$ or $V_t$ in Fig.~\ref{fig:fscan_all}), we can map $\chi$ as a function of $n_g$ (Fig.~\ref{fig:charge_scan}).  For the measured devices, we observe a maximum $2\chi_0$ of $\sim$10\,MHz, about ten times the linewidth.  These charge scans provide direct verification of parity-state tunability and allow extraction of $E_J$ and $E_C$ from the combined dependence of $f_0$ and $\chi_0$ (see Eq.~\ref{eqs:squatParams}).

As optimal readout fidelity occurs when $\chi$ satisfies Eq.~\ref{eq:dispersion}, one may use this DC bias voltage to maintain a charge state for the duration of parity measurements. The voltage required to shift the offset charge by a full $2e$ period depends on the capacitive coupling strength of the bias line. On large multi-device chips, this dependence provides a potential handle for device identification and discrimination.

\subsection{Pulsed Measurements} \label{sec:pulsed_meas}

SQUAT qubit properties can be characterized by sending coherent microwave pulses through the transmission line and measuring the subsequent emission.  In this section, we describe pulsed measurements used to investigate the Rabi rate~($\Omega$), the energy relaxation time~($T_1$), and the dephasing time~($T_2^*$).  The measurements are conducted with the SQUAT biased to the charge degenerate point, such that only one resonance appears and $f_r$ remains constant with parity changes.  We use high-power drive pulses such that $\Omega \gg \chi$ and we can approximate the qubit as a two-level system.

\subsubsection{Rabi Measurements}

\begin{figure}
    \centering
    \includegraphics[width=1.0\columnwidth]{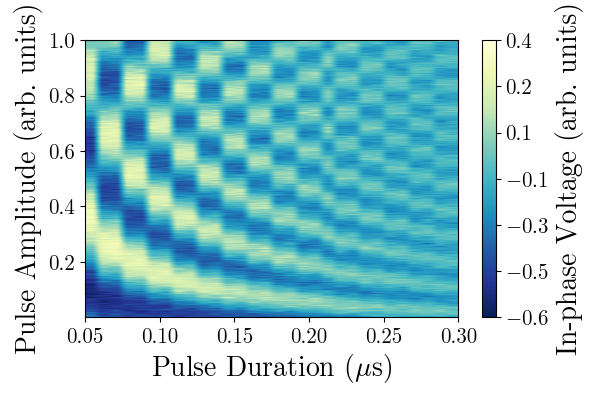}
    \caption{Rabi oscillations in emitted field plotted as a function of drive pulse duration and amplitude.  The pulse frequency is matched to $f_0$, and each point consists of 50,000 averages.}
    \label{fig:rabi}
\end{figure}

To measure Rabi rate, we apply a drive pulse at the qubit transition frequency.  The pulse causes the qubit to precess in the Bloch sphere (see App.~\ref{app:pulsed_ro_diagrams}), as governed by the Hamiltonian (see App.~\ref{app:hamiltonian}).  After each pulse, we turn off the drive tone and measure the emitted field during a short averaging time (typically 5\,ns).  The amplitude and phase of the post-pulse emission oscillate as a function of pulse duration and amplitude, thus tracing out coherent Rabi oscillations (Fig.~\ref{fig:rabi}). The power dependence of $\Omega$ can be described using readout power or photon interrogation rate:

\begin{equation}\label{eq:PeropadreEquation}
    \frac{\Omega^2}{\Gamma_r} = 2\Gamma_n = \frac{4 P_r}{hf}
\end{equation}

This relation enables direct comparison to resonance fits from Eq.~\ref{eq:S21Squat} taken at equal drive power.  The Rabi measurement is also used to calibrate $\pi$ and $\pi/2$ pulses.

Unlike resonator-coupled transmons, which observe Rabi oscillations in measurements of qubit excitation ($\langle\sigma_z \rangle$), SQUAT emission measures the expectation value of the de-excitation operator, $\langle \sigma_- \rangle =  ( \langle \sigma_x \rangle - i \langle \sigma_y \rangle ) /2 $. A null signal corresponds to the qubit occupying a pure energy eigenstate, where no coherent radiation is emitted. A peak in emission corresponds to a state in the XY plane of the Bloch sphere. Additionally, measured Rabi oscillations appear with a sinusoidal envelope determined by the detuning ($\delta f$) between the pulse and qubit frequencies (see App.~\ref{app:pulsed_ro_diagrams} for additional discussion).

\subsubsection{Dephasing Measurement ($T_2^*$)}

To measure the qubit dephasing time including low-frequency noise effects ($T_2^*$), we first prepare the qubit in the $|+\rangle$ state using a $\pi/2$ pulse.  After the pulse, we wait a variable delay time ($\tau$), then trigger the ADCs to measure spontaneous emission to the feedline. During this time, the qubit precesses under any detuning ($\delta f$) between the drive and qubit frequency. 

The emitted field is proportional to $\langle \sigma_-(\tau) \rangle$, which evolves as $\langle \sigma_-(\tau) \rangle \propto e^{-\tau/T_2^*} e^{-i 2\pi \delta f \tau}$.  The measured real amplitude is therefore
\begin{equation}
A(\tau) \propto e^{-\tau/T_2^*} \cos(2\pi\delta f \tau + \phi_0),
\end{equation}
where $\phi_0$ is a fixed phase offset. By varying $\tau$, we record the decaying oscillations and fit to extract $T_2^*$ (Tab.~\ref{tab:fitValues}).  Off-resonance, the oscillation reflects XY-plane precession, while the envelope gives the dephasing.

\subsubsection{Energy Relaxation ($T_1$)}

Measuring qubit energy relaxation time ($T_1$) in this basis requires a two-pulse sequence.  We begin by exciting the qubit to the $|1\rangle$ state using a $\pi$ pulse.  We then wait a variable time $\tau$, during which the qubit may relax.  A $\pi/2$ pulse is then applied to rotate the qubit back into the XY plane.  After this second pulse, we trigger our ADCs for a short integration and plot signal amplitude.  If the qubit maintained its state for the duration of the wait time, it is mapped to the $|-\rangle$ state.  If the qubit decayed to $|0\rangle$ during the delay time, the $\pi/2$ pulse will map it to the $|+\rangle$ state, leading to opposite signal amplitude in IQ-space: 
\begin{equation}
    \langle \sigma_- (t) \rangle \propto 
    \begin{cases}
    +1 & \text{if qubit decayed to } |0\rangle \\
    -1 & \text{if qubit remained in } |1\rangle
    \end{cases}
\end{equation}
After repeated measurements, we express the excitation probability as a function of $\tau$
\begin{equation}
    P_1 (\tau) = \frac{1 - \langle \sigma_- (\tau) \rangle}{2}
\end{equation}
We then fit the excited population to an exponential
\begin{equation}
    {P}_1 (\tau) = e^{-\tau/T_1}
\end{equation}
to extract the qubit's energy relaxation time, $T_1$ (Tab.~\ref{tab:fitValues}).


\begin{table*}[t]
    \centering
    \caption{SQUAT parameters for the five devices discussed in this paper. $f_0$ and $2\chi_0$ are extracted from frequency-dependent transmission measurements taken with a VNA. $E_J$ and $E_C$ are subsequently calculated using Eqs.~\ref{eqs:squatParams}.  The decay rate $\Gamma$'s were found using fits to Eq.~\ref{eq:S21Squat}. For each device, $2\gamma$ is dominated by $\Gamma_r$, which is dominated by $\Gamma_c$. Decoherence times calculated by inverting $\gamma$ and $\Gamma_r$ are included in their respective columns (see App.~\ref{app:hamiltonian}). The decoherence times reported in the last two columns were independently found using pulsed measurement sequences, explained in Sec.~\ref{sec:pulsed_meas}.  There is reasonable agreement between $T_1$ from the VNA and pulsed measurements.  However, we see suppression in $T_2^*$ from pulsed measurements compared to $T_2$ from the VNA.  This is attributed to the low-frequency systematics that contribute to $T_2^*$ measurements, as well as sensitivity to charge noise being maximized when the qubit is biased to $\chi=0$ for pulsed measurements.}
    \label{tab:fitValues}
    \begin{tabular}{|c|c|c|c|c|c|c|c|c|c|}
    \hline
        Device & $f_0$ (GHz) & $2\chi_0$ (MHz) & $E_J/h$ (GHz) & $E_C/h$ (MHz) & $\gamma$ (MHz) & $\Gamma_r$ (MHz) & $\Gamma_c$ (MHz) & $T_2^*$ (ns) & $T_1$ (ns) \\
        & & & & & $ \rightarrow T_2$ (ns) & $\rightarrow  T_1$ (ns) & & & \\
        \hline
         H1Q1 & 7.35 & 9.46 & 12.8 & 626 & $3.670 \pm 0.003$ & $6.35 \pm 0.06$ & $6.301 \pm 0.004$ & $291 \pm 18$ & $210 \pm 8$ \\
          & & & & &  $  \mathit{\rightarrow 266.5 \pm 0.2} $ & $\mathit{\rightarrow 156 \pm 1}$ & & & \\
         H1Q2 & 8.18 & 3.10 & 15.7 & 622 & $5.564 \pm 0.002$ & $11.106 \pm 0.005$ & $11.140 \pm 0.004$ & $143 \pm 5$ & $ 98 \pm 3$  \\
         & & & & & $\mathit{\rightarrow 179.54 \pm 0.08}$ & $\mathit{\rightarrow 90.00 \pm 0.04}$ & & & \\
         H2Q1 & 8.37 & 1.92 & 16.6 & 615 & $10.01 \pm 0.09$ & $19.3 \pm 0.1$ & $18.8 \pm 0.1$ & $116 \pm 2$ & $81 \pm 2$ \\
         & & & & & $\mathit{\rightarrow 98.8 \pm 0.9}$ & $\mathit{\rightarrow 51 \pm 3}$ & & & \\
         H2Q2 & 8.42 & 2.11 & 16.7 & 618 & $4.80 \pm 0.03$ & $9.60 \pm 0.06$ & $9.61 \pm 0.06 $ & $106 \pm 5$ & $103 \pm 9$ \\
         & & & & & $\mathit{\rightarrow 207 \pm 1}$ & $\mathit{\rightarrow 103.7 \pm 0.7}$ & & & \\
         H2Q3 & 8.47 & 1.57 & 17.2 & 606 & $5.6 \pm 0.1$ & $10.7 \pm 0.7$ & $10.4 \pm 0.2$ & $85 \pm 3$ & $104 \pm 4$ \\
         & & & & & $\mathit{\rightarrow 174 \pm 4}$ & $\mathit{\rightarrow 93 \pm 6}$ & & & \\
         \hline
    \end{tabular}
\end{table*}

\section{Parity Measurements}\label{sec:parity_meas}

The core functionality of the SQUAT sensor is its ability to resolve parity-switching events caused by individual quasiparticle tunneling.  As a result of its direct feedline coupling, the SQUAT parity state can be monitored using a single CW tone.  We do so in two different operating modes distinguished by the tone location relative to the even/odd parity bands.

`Amplitude readout' places the tone on resonance with one parity band, so parity-switching events appear as jumps in transmitted amplitude.  In this readout mode, the signal-to-noise ratio~(SNR) can be enhanced by a factor of $\sqrt{2}$ by simultaneously driving both parity bands and taking the difference of the two signals.  This is the standard benefit from combining two measurements with uncorrelated noise.

`Phase readout' places the tone at the midpoint between the two parity bands, where parity-switching events cause a change in the phase of the transmission. When optimized for power and dispersion, the intrinsic SNR of phase readout is twice that of amplitude readout (see App.~\ref{app:fidelity}) without requiring multiple tones.  The phase difference between parity bands depends on $\chi$, which depends on global offset charge ($n_g$), allowing phase readout to measure both charge and parity signals simultaneously.  This readout mode also eliminates the need for tone-tracking in response to changes in $n_g$.

\begin{figure}
    \centering
    \includegraphics[width=1.0\columnwidth]{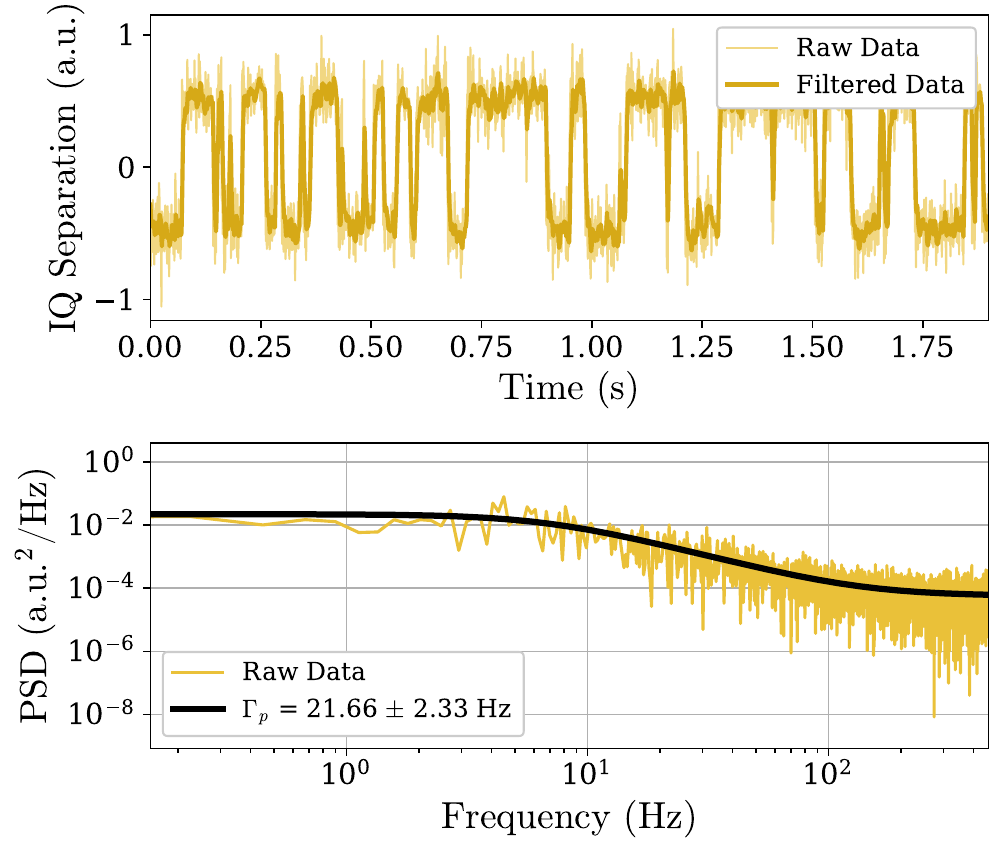}
    \caption{A representative parity-switching measurement. \textbf{Top:} Raw and filtered data plotted as a function of time. The data is a small slice of a longer time-domain acquisition and has been rotated to the basis of maximum signal.  \textbf{Bottom:} PSD of the full dataset with a fit to Eq.~\ref{eq:fidelity_main} to extract the characteristic switching rate.} 
    \label{fig:timestream}
\end{figure}

When complex transmission is recorded in either mode, the telegraph signal produces two Gaussian clusters in IQ space, corresponding to the even and odd parity states.  The separation of the two clusters (and thus signal amplitude) will depend on $\chi$ and $\Gamma_n$.  A discussion of readout-power optimization is included in App.~\ref{app:fidelity}.  The clusters are fit to find the axis of maximum separation, and the projection of the IQ data onto this axis is used for analysis.  Fig.~\ref{fig:timestream} shows the telegraph signal in the maximal separation basis for phase readout, where parity-switching is clearly visible in both the raw and Butterworth-filtered datasets.

The readout fidelity~($\mathcal{F}$) is related to the projected-state variance (Eq.~\ref{eq:stateVariance_main}) via 
\begin{equation}
    \mathcal{F} = \mathrm{erf}\left(\frac{1}{\sqrt{8\sigma_s^2}}\right) .
\end{equation}
This fidelity introduces a noise floor to the power spectral density (PSD) of the parity timestream. The PSD can be fit with the following spectral form to extract both $\mathcal{F}$ and the characteristic parity-switching rate~($\Gamma_{p}$) as a function of sampling bandwidth ($f_{bw}$)~\cite{Rist__2013}.
\begin{equation}\label{eq:fidelity_main}
    \text{PSD}(f) = \mathcal{F}^2\frac{4\Gamma_{p}}{(2\Gamma_{p})^2+(2\pi f)^2} + (1-\mathcal{F}^2)f_{bw}^{-1}
\end{equation}
App.~\ref{app:parity} presents a Monte Carlo validation of the relation between variance and fidelity for a given effective noise temperature and $f_{bw}$.

The switching shown in Fig.~\ref{fig:timestream} is $\sim20$\,Hz background tunneling due to the device's environment and residual quasiparticle density.  The rate of background switching can be calculated either by filtering time-domain data and counting events per unit time, or by fitting the PSD to Eq.~\ref{eq:fidelity_main}.  We used the latter method to calculate all $\Gamma_{p}$ values in this paper because it did not require filter parameter adjustments for individual datasets.

For the SQUATs measured in this paper, we find a noise temperature around 10.37\,K, which gives a fidelity of 50\% at sampling bandwidth of 6\,kHz (see App.~\ref{app:fidelity} for more details). When accounting for the symmetry of our waveguide, we actually expect the SQUAT itself to emit 50\% of the photons back to the input line, leading to an output-referred noise temperature of 5.19\,K. Compared to the expected HEMT noise temperature (1.8\,K), this implies a loss of around 5\,dB between the SQUAT and amplifier input, which is reasonable considering the filters, interconnects, and circulators used between the device and amplifier.

\subsection{Temperature Dependence} \label{sec:temp_dep}

\begin{figure}
    \includegraphics[width=1.0\columnwidth]{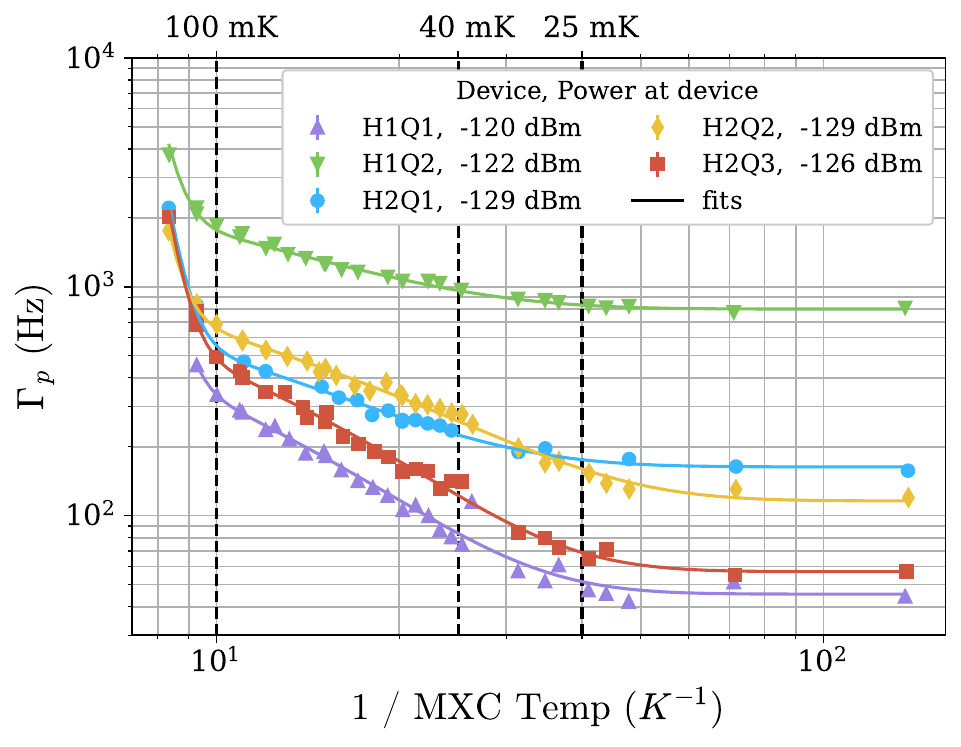}
    \caption{Parity-switching rate as a function of mixing-chamber temperature, plotted for five SQUATs on chips `H1' and `H2'. Different readout powers were chosen to maintain sufficient signal-to-noise for each SQUAT. Dashed vertical lines delineate temperature regions of interest.}
    \label{fig:parityTempSwitching}
\end{figure}

The background parity-switching rate depends on the residual non-equilibrium quasiparticle density~($x_\text{qp}^\text{ne}$) and the tunneling barrier~($\delta \Delta$) caused by the difference in superconducting gap between the two junction leads.  To characterize these parameters, we measure $\Gamma_p$ as a function of mixing-chamber temperature.  We use a PID loop to heat the mixing chamber and allow the device to thermalize at each temperature before measuring $\Gamma_p$.  The resulting behavior, shown in Fig.~\ref{fig:parityTempSwitching}, exhibits three distinct temperature regimes.  At low temperatures (below $\sim$25\,mK), the switching rate is essentially flat, demonstrating dominance of non-thermal backgrounds such as photon-assisted processes~\cite{Diamond_2022, Connolly2024}.  In the intermediate temperature regime (25 to 100\,mK), we see thermal activation of the rate, as the increased temperature assists ambient quasiparticles in overcoming the tunneling barrier.  At the highest temperatures~($>$100\,mK), the quasiparticle density increases, leading to an exponential increase in parity-switching rate.

Following Ref.~\cite{Nho_2025}, we fit our data using a model that combines these effects and accounts for the dependence of rate on qubit excitation probability~($P_1$). With $h f_0 > \delta\Delta$ (as is the situation here), de-excitation of the qubit can contribute sufficient energy to induce tunneling of an existing quasiparticle. Therefore, higher $P_1$ leads to higher switching rate.  Since CW monitoring of the device parity drives the qubit into a mixed state with a power-dependent $P_1$, we expect the rate to be a linear combination of the excited~($\Gamma_1$) and non-excited~($\Gamma_0$) switching rates. We also include a temperature-independent background~($\Gamma_{\text{other}}$) with no assumption of its origin.  The overall switching rate is thus
\begin{equation} \label{eq:gamma_sum}
    \Gamma_{p}(1/T)= (1-P_1) \Gamma_0 + P_1\Gamma_1 + \Gamma_{\text{other}}.
\end{equation}

When reading out in `phase readout mode' with low power ($2\Gamma_n/\gamma \ll 1$), we use Eq.~\ref{eq:P1} to find $P_1\approx 0$ and $\Gamma_p$ to be dominated by $\Gamma_0$ and $\Gamma_\text{other}$.  For intuition building, we consider $\Gamma_0$ in the limit $\Delta \gg \delta \Delta$ and $hf_0 \gg k_B T$.
\begin{equation} \label{eqs:Gamma_0}
    \Gamma_0(T) = \frac{ x_\text{qp}^\text{ne} }{1+\frac{V_H}{V_L}e^{\frac{-\delta \Delta}{k_B T}}} \frac{16 E_J \delta \Delta}{h \sqrt{2 \pi k_B T \Delta}} K_1 \left( \frac{\delta \Delta}{2 k_B T} \right) e^{\frac{-\delta \Delta}{2 k_B T}}
\end{equation}
where $T$ is the device temperature and $V_H$ ($V_L$) is the volume of the high (low) gap sides of the junction.  In this formulation, the two junction leads have superconducting gaps $\Delta$ and $\Delta + \delta \Delta$.  $K_1$ is the first-order modified Bessel function of the second kind.

At the lowest temperatures ($2k_b T \ll \delta \Delta $), the equation reduces to
\begin{equation} \label{eqs:Gamma_0_limit}
    \Gamma_0 (T\rightarrow0) = x_\text{qp}^\text{ne} \frac{16 E_J}{h} \sqrt{\frac{\delta \Delta}{2 \Delta}} e^{\frac{-\delta \Delta}{ k_B T}}.
\end{equation}
The $\Gamma_0$ contribution is exponentially suppressed, so $\Gamma_p$ is dominated by $\Gamma_{\text{other}}$, as seen in Fig.~\ref{fig:parityTempSwitching}.

For SQUAT frequencies around 8\,GHz, the assumption $hf_0 \gg k_B T$ is only valid at the lowest temperatures~($\lesssim40$\,mK).  We fit the full temperature scan to the exact equation for $\Gamma_0$ using App.~\uppercase\expandafter{\romannumeral2\relax}b of Ref.~\cite{Nho_2025}.  The results of these fits are detailed in Tab.~\ref{tab:temp_fit}, and give $x_\text{qp}^\text{ne} \sim \mathcal{O}(10^{-8})$ and $\delta \Delta\approx2$\,GHz, which matches expectations from fabrication. We also use the fitted $\Delta$ to estimate the transition temperature of our Al film to be $\sim1.1$\,K.

\begin{table}[]
    \centering
    \caption{Resulting parameters from the temperature-dependence fits in Fig.~\ref{fig:parityTempSwitching}.}
    \label{tab:temp_fit}
    \begin{tabular}{|c|c|c|c|c|}
    \hline
        Device & $x_\text{qp}^\text{ne}$ $(\times10^{-8})$ & $\delta \Delta$ (GHz) & $T_c$ (K) & $\Gamma_{\text{other}}$ (Hz)\\
        \hline
         H1Q1 & $1.84\pm0.05$ & $2.49\pm0.05$ & $1.126\pm0.009$ & $45.4\pm0.8$   \\
         H1Q2 & $4.6\pm0.2$ & $2.2\pm0.1$ & $1.082\pm0.005$ & $799\pm8$   \\
         H2Q1 & $1.65\pm0.10$ & $2.2\pm0.1$ & $1.102\pm0.004$ & $163\pm2$   \\
         H2Q2 & $2.11\pm0.04$ & $1.57\pm0.03$ & $1.146\pm0.003$ & $116\pm2$   \\
         H2Q3 & $1.89\pm0.04$ & $2.29\pm0.04$ & $1.107\pm0.002$ & $56.9\pm1.0$   \\
         \hline
    \end{tabular}
\end{table}

\subsection{Background-Rate Drivers}

In the devices studied, the lowest background rate was found to be under 10\,Hz and dominated by $\Gamma_{\text{other}}$. To better understand the drivers of background rate, we performed a systematic study of various effects and discuss the resulting trends below.

\subsubsection{IR environment}\label{sec:IR_env}

Infrared~(IR) black-body radiation from higher-temperature stages is a known source of quasiparticle poisoning in transmons (see Refs.~\cite{Liu2022,Connolly2024}).  Likewise, when mounted in a non-light-tight magnetic shield, our SQUAT devices exhibited parity-switching rates too high to resolve~($>$10\,kHz). 

To reduce the population of IR-generated quasiparticles, we adopted an enclosure design based on Ref.~\cite{Iaia_2022}.  The enclosure uses a Cryoperm can coated internally with black IR absorber and sealed by a light-tight (no direct optical path) copper lid that incorporates IR stub filters~\cite{yyStubFilter}.  With the new shielding configuration, the switching rate was reduced by multiple orders of magnitude, enabling the measurements presented above.  Additional details on the enclosure are included in App.~\ref{app:device_env}.

\begin{figure}
    \includegraphics[width=1.0\columnwidth]{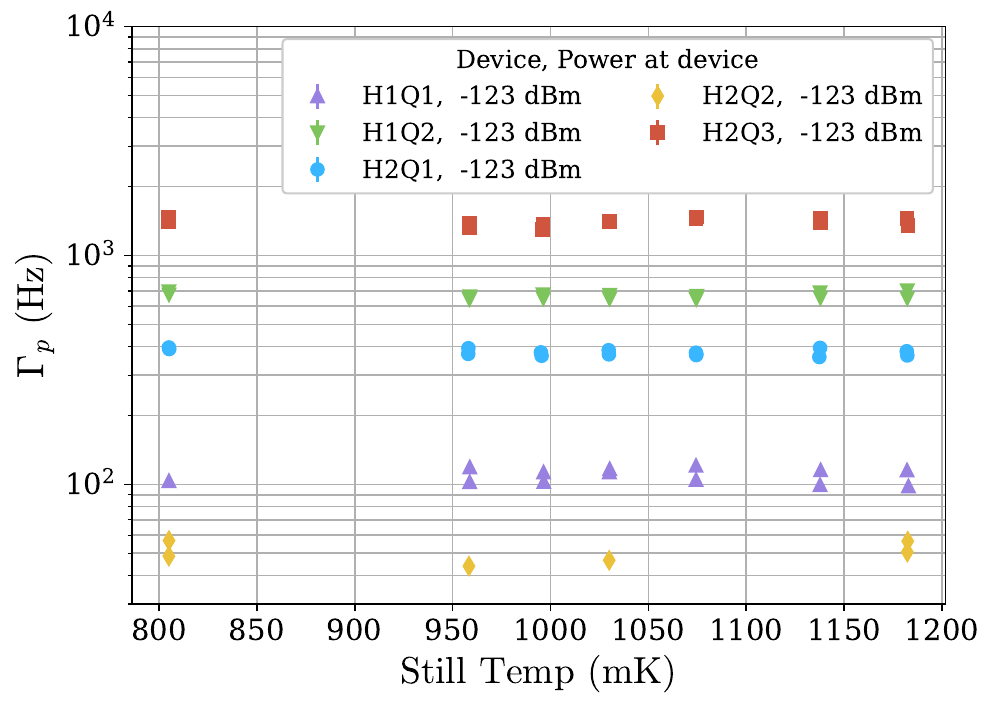}
    \caption{Parity-switching rate as a function of still plate temperature, controlled using a heater. For the duration of this dataset, the mixing chamber temperature was held at $20.00\pm0.06$\,mK.  We note that this measurement was taken during a different fridge run than Fig.~\ref{fig:parityTempSwitching} and displays rate variation discussed in Sec.~\ref{sec:IR_env}}.
    \label{fig:StillTempSweep}
\end{figure}

The effectiveness of this shielding was tested by heating the still stage of the dilution refrigerator while maintaining constant mixing-chamber temperature, thereby modifying the black-body environment without changing device performance. The result showed no measurable dependence of switching rate on still temperature, indicating that IR leakage from the still is a subdominant source of quasiparticle poisoning (see Fig.~\ref{fig:StillTempSweep}).

Variation in parity-switching rate between qubits and fridge cooldowns suggests a non-intrinsic background driving device performance. A potential explanation is IR radiation reaching the detectors through the coaxial connections to each device.  Our measurement setup employs comparable coaxial IR filtering to one used in Ref.~\cite{Nho_2025}, and we measure a similar rate (see App.~\ref{app:device_env} for more details). In Ref.~\cite{Nho_2025}, parity-switching rate was further reduced with additional filtering, which we will attempt to reproduce in future work.

\subsubsection{Readout Power}

For SQUATs with switching rates below $\sim200$\,Hz, we observed an increase in switching rate with CW readout power (see App.~\ref{app:readout_power}).  This power-induced background has several proposed explanations which may work in combination.  First, higher drive power increases $P_1$, which enhances the rate through de-excitation-assisted tunneling~\cite{Connolly2024}.  Second, increased dissipation in attenuators and other components can produce a black-body tail of pair-breaking photons.  Finally, since the readout photons have energy $hf_0 > \delta\Delta$, it is possible that photons from the transmission line can assist ambient quasiparticles in overcoming the energy barrier and tunneling.  These proposed explanations will be further decoupled in future studies.

\subsubsection{Vibrations}

Prior work indicates that pulse-tube-induced vibrations can contribute noise to phonon-sensitive devices such as SQUATs~\cite{yelton2025}.  Consequently, measurements with pulse tubes turned off are standard in our characterization protocol.  Baseline parity-switching measurements for two SQUATs in separate cryostats are reported in Tab.~\ref{tab:PTswitch} (also see App.~\ref{app:vibration}).  In both tests, the pulse tubes were turned off, and data was immediately taken for approximately two minutes, resulting in a decreased switching rate.  In each setup, the relative contribution of pulse-tube induced tunneling may depend on a variety of factors, including fridge design, device mounting, and ambient quasiparticle density. Vibration mitigation for optimal device performance is an ongoing study.

\begin{table}[]
    \centering
    \caption{Comparison of switching rates for SQUAT devices between two different cryostats with identical wiring, with the pulse tube on and off.}
    \label{tab:PTswitch}
    \begin{tabular}{|c|c|c|}
    \hline
       Cryostat & PT On & PT Off \\
    \hline
        ProteoxMX & $250\pm30$ \,Hz & $38\pm9$\,Hz \\
        BlueFors XLD400 & $43\pm6$\,Hz & $19\pm4$\,Hz \\
    \hline
    \end{tabular}
\end{table}

\section{Discussion}\label{sec:discussion}

\begin{figure}
    \centering
    \includegraphics[width=1.0\columnwidth]{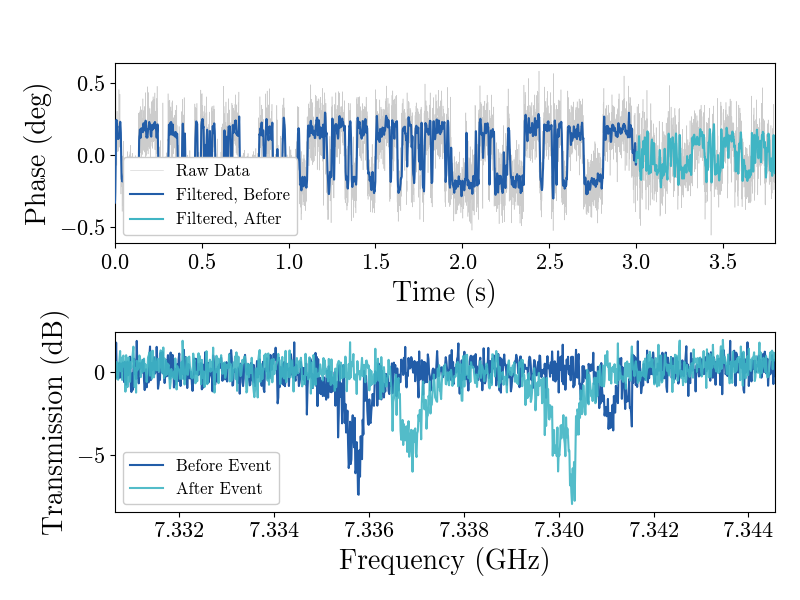}
    \caption{An example of a charge-triggered event which occurred in the device substrate, detected via the `Phase Readout' mode described in Sec.~\ref{sec:parity_meas}. The event produced free charge, which induced an offset charge on the SQUAT and contributed to an elevated parity-switching rate. The offset charge jump can be seen by the change in separation of the parity bands, reflected in VNA scans taken before and after the time-sequence. The parity-switching trace averages to zero following the event, indicating that switching is occurring more rapidly than our readout bandwidth.}
    \label{fig:chargeEvent}
\end{figure}

The results presented here constitute the first experimental demonstration of a SQUAT-style sensor, establishing a new platform for studying single-quasiparticle dynamics and high-efficiency detection of quasiparticle-generating events.  Unlike conventional transmon devices, which are typically engineered to suppress quasiparticle sensitivity, the SQUAT geometry allows for high-fidelity monitoring of tunneling events with CW readout.

The present work serves to establish baseline detector performance and readout requirements, provide a complete set of measurements necessary for new device characterization, and identify major background sources (including IR leakage, readout-power-assisted switching, and vibration-induced phonons). Future measurements of existing and novel SQUAT devices will focus on disentangling and suppressing these backgrounds.

In an initial validation of detector-style operation, we also observe the coincident occurrence of switching-rate bursts and global offset-charge shifts, which can be read out simultaneously with a single CW tone (as shown in Fig.~\ref{fig:chargeEvent}). Future correlation measurements of charge-parity coincidence could serve either as a veto of high-energy events, or as a probe of charge and phonon propagation across multi-SQUAT chips.

With this architecture validated, next-generation detectors are being fabricated with gap-engineered quasiparticle trapping in the region of the junction. This effort includes designs with standard Al junctions and higher-Tc collection islands as well as Al islands (with long quasiparticle diffusion lengths) and lower-Tc junctions (using materials such as Hf~\cite{balaji2025}).  These devices are expected to have improved event-detection efficiency~\cite{fink_PD2}.

To characterize the energy sensitivity of these and future devices, upcoming studies will use controlled energy depositions generated with optical LEDs (Refs.~\cite{Stifter_2025, tabassum2025_MEMS_chopping}), THz photo-mixers, or radioactive sources.  Time-domain analysis of the SQUAT response will enable determination of an absolute detector energy calibration, phonon collection efficiency, and THz antenna coupling.  These measurements will be crucial in the development of next-generation detectors targeting meV phonon and THz photon sensitivities.

Lastly, the intrinsic compatibility of SQUATs with frequency-domain multiplexing provides a path towards large arrays with improved active-sensor coverage.  These multi-SQUAT devices will provide an experimental platform for coincidence studies, spatial-dependence measurements, and low-background event searches.

Forthcoming work will build heavily on the results presented here, which provide a basic confirmation that the design proposed in Ref.~\cite{fink2024} behaves as anticipated. Such work will focus on building devices more tailored to specific phonon and photon sensing applications, developing robust microwave-multiplexed readout, and studying the quasiparticle trapping and amplification effects necessary to maximize SQUAT sensitivity.  SQUATs are proving to be an effective tracer of environmental effects observed in qubits and continue to show promise as quantum sensors for discovery science.   

\begin{acknowledgments}
This work was supported in part by the US Department of Energy Early Career Research Program (ECRP) under FWP 100872. H. Magoon was supported by a Kavli Institute for Particle Astrophysics and Cosmology graduate student fellowship, and by NSF GRFP.  C. Fink was supported by the New York State Empire State Development Agency grant \#134639. C. P. Salemi was supported by the Kavli Institute for Particle Astrophysics and Cosmology Porat Fellowship. B. A. Young was supported in part by a Lee and Seymour Graff Endowed Chair grant. This work was supported in part by the DOE Office of Science High Energy Physics QuantISED program. This work was supported in part by a Graduate Research Instrumentation Award (GIRA) through the Coordination Panel for Advanced Detectors (CPAD) as funded by the  DOE Office of High Energy Physics. This material is based upon work supported by the U.S. Department of Energy Office of Science National Quantum Information Science Research Centers as part of the Q-NEXT center. Q-NEXT supported engineering and analysis effort for an early stage of device testing and is supporting ongoing work in photon-coupled SQUAT readout at GHz-THz frequencies.  Device fabrication was performed at nano@stanford RRID:SCR\_026695.

The authors would like to thank Ziqian Li and Fanghui (Wendy) Wan for fabrication training and recipe development, as well as Amir Safavi-Naeini's group for providing access to their Plassys evaporator.  We thank Matthew Maksymowych in particular for maintaining the tool at the time of this fabrication.  We also thank Shawn Henderson for guidance on device readout, as well as cryogenic and facility support. Finally, the authors would like to thank Britton Plourde and Robert McDermott for discussion and advice in the early stages of fabrication and measurement, and for guidance related to photon poisoning of qubits via antenna modes.
\end{acknowledgments}

\bibliographystyle{aipnum4-1}
\bibliography{qubit}

\clearpage
\pagebreak
\onecolumngrid
\appendix

\section{Variables Summary} \label{app:params}

To help with readability, we include here a list of parameter definitions. Note that in the main text, we use frequency for clarity in comparison to measurement parameters, whereas in some of the appendix we retain angular frequency for conciseness.

\begin{table}[htbp]
    \centering
    \caption{List of Variables}
    \label{tab:variables}
    \begin{tabular}{|c|p{13cm}|}
        \hline
        \textbf{Variable} & \textbf{Description} \\
        \hline
        $E_J$ & Josephson energy of the SQUAT junction ($=\Phi_0 I_c/2\pi$, with $I_c$ being the junction critical current)\\
        $E_C$ & Charging energy of the SQUAT, $E_C = e^2/2C_\Sigma$, with $C_\Sigma$ being the total island capacitance)\\
        $2\chi$ & Charge-dependent frequency separation between even and odd parity states \\
        $2\chi_0$ & Maximum charge-dependent frequency separation between even and odd parity states \\
        $f$ & Applied microwave drive (or probe) frequency\\
        $f_0$ & Bare qubit transition frequency in the absence of charge dispersion \\
        $f_r$ & Instantaneous qubit transition frequency, including charge-dependent dispersion and parity state ($f_r = f_0 \pm \chi$ ) \\
        $\delta f$ & Detuning between the applied drive and qubit transition frequency ($\delta f = f - f_r$)\\
        $n_g$ & Dimensionless offset charge on the qubit island, expressed in units of Cooper-pair charge $2e$\\

        $P_{\text{VNA}}$ & Output power of the vector network analyzer (VNA) at the instrument source\\        
        $P_r$ & Readout power incident on the device \\ 
        $\Gamma_n$ & Photon interrogation rate ($\Gamma_n = 2 P_r/hf_r$) \\

        $\gamma$ & Total qubit decoherence rate ($ \gamma =\Gamma_r/2+\Gamma_{\phi}$)\\
        $\Gamma_r$ & Total radiative decay rate of the qubit ($\Gamma_r=\Gamma_c + \Gamma_l$) \\
        $\Gamma_c$ & Radiative decay rate into the transmission line \\
        $\Gamma_l$ & Radiative decay rate into non-transmission line loss channels \\
        $\Gamma_\phi$ & Pure dephasing rate, excluding energy relaxation \\

        $Q_r$ & Total qubit quality factor \\
        $Q_c$ & Coupling quality factor, associated with decay to the transmission line\\
        $Q_i$ & Internal quality factor, associated with intrinsic loss mechanisms\\

        $\sigma_s^2$ & Variance of the inferred qubit state, inversely proportional to the signal-to-noise ratio ($=\text{SNR}^{-1}$)\\
        $\sigma_V^2$ &  Variance of voltage noise in the readout chain\\
        $f_{bw}$ & Effective measurement bandwidth of the readout \\
        
        $T_2^*$ & Qubit dephasing time, including contributions from low-frequency noise \\
        $T_2$ & Qubit dephasing time, excluding contributions from low-frequency noise \\
        $T_1$   & Qubit energy relaxation time \\
        $T_\phi$ & Pure dephasing time, characterizing loss of phase coherence in the absence of energy relaxation ($1/T_2 = 1/(2T_1) + 1/T_\phi$)\\
        
        $\tau$ & Variable time delay used in pulsed measurement sequences \\ 
        $P_1$ & Probability of the qubit being in the excited state \\
        $\Omega$ & Rabi rate (angular frequency) \\
        $\mathcal{F}$ & Readout fidelity \\ 
        $\Gamma_{p}$ & Total parity-switching rate\\
        $\Gamma_0$ & Parity-switching rate when the qubit is in the ground-state \\
        $\Gamma_1$ & Parity-switching rate when the qubit is in the excited-state \\
        $\Gamma_{\text{other}}$ & Rate of parity-switching due to unspecified temperature-independent backgrounds \\ 
        $\Delta$ & Superconducting energy gap\\
        $\delta \Delta$ & Difference in superconducting gap between junction leads\\
        $T_c$ & Superconducting critical temperature \\
        $x_\text{qp}^\text{ne}$ & Dimensionless residual non-equilibrium quasiparticle density, normalized to the Cooper-pair density \\
        $V_H$, $V_L$ & Volumes of the high ($V_H$) and low ($V_L$) gap sides of the junction \\ 
        $T_n$ & Noise temperature \\
        $T_\text{eff}$ & Effective noise temperature \\
        \hline
    \end{tabular}
\end{table}

\section{SQUAT Design and Simulation}\label{app:design}

A three-SQUAT prototype was designed to demonstrate multiplexed, direct-coupled readout without a resonator, while keeping device parameters within a practical range for fast, high-fidelity measurements.  Target design values included quality factors of $Q_r \approx 1000$, yielding a readout bandwidth of $f_{bw} \approx f_0/Q_r \approx 1$ MHz. Maximum charge dispersion was chosen to be $2\chi_0 \approx 10$\,MHz, which provides clear parity-state separation under the optimal readout condition introduced in the main text.  To minimize crosstalk and unintended couplings, the central frequency of the three qubits were spaced by roughly 50\,MHz.  The designs were nominally identical, with small geometric adjustments to set $f_0$ and $\chi_0$.  The resulting simulated parameters and key dimensions for devices A, B, and C are summarized in Tab.~\ref{tab:designValues}.

Each SQUAT uses a bowtie-shaped capacitor geometry to enhance photon coupling and promote quasiparticle diffusion towards the junction.  The island capacitance, and consequently the charging energy $E_C$, is set by the size and placement of the capacitor pads. All SQUATs are coupled to a common coplanar feedline. Each is oriented with one capacitor island closer to the feedline than the other, enabling differential bias of all devices through the feedline. An auxiliary, weakly-coupled bias line on the opposite side of each SQUAT allows for independent charge tuning without introducing a significant non-readout decay channel.

The chip is designed with a niobium (Nb) ground plane and a coplanar waveguide feedline shared by all three SQUATs. Nb was chosen for its high superconducting gap, making phonons absorbed in the ground plane or waveguide less likely to produce quasiparticles and more likely to return to the substrate, where they can be absorbed by the lower-gap aluminum (Al) islands. Al has a high quasiparticle-diffusion length, improving the fraction of island quasiparticles that reach the junction before recombining.  Both ground plane and Al islands include flux-trap hole arrays with 55\,$\mu$m spacing to mitigate vortex-related loss.  A schematic of one SQUAT and the full chip layout are shown in Figs.~\ref{fig:designOverview} and \ref{fig:devicePackage}.  The junctions were deposited with a standard Al/AlOx/Al recipe, and were designed with asymmetric lead thicknesses (45\,nm and 115\,nm).  This produces a small gap asymmetry ($\delta\Delta$), which can be measured with temperature scan datasets (detailed in Sec.~\ref{sec:parity_meas}).  Future work will continue to explore the trade-off between quasiparticle tunneling efficiency and background parity-switching rates. 

Electromagnetic simulations informed the device geometry.  Each SQUAT was simulated independently in HFSS (Driven Modal), with the Josephson junction modeled as a lumped inductor $L_J \approx 10\,$nH  consistent with the assumed  $E_J/h \approx 16.3\,$GHz:
\begin{equation}
L_J = \left(\frac{\Phi_0}{2\pi}\right)^2 \frac{1}{E_J} \approx 10\,\mathrm{nH}\left[\frac{16.3\,\mathrm{GHz}}{E_J/h}\right].
\end{equation}
HFSS provided $f_0$ and the resonance quality factor $Q_r$. $E_C$ was then estimated using the standard transmon relations given $E_J$ and $f_0$ (see Eq.~\ref{eqs:squatParams}).  The capacitance matrix was extracted from Ansys Maxwell electrostatic simulations in order to cross-check $E_C$ and confirm that charge-line coupling is sufficient to DC bias the devices with reasonable on-chip voltages.  The simulated $E_C$ values showed slight deviations from measurements and device-to-device variation consistent with micron-scale variability in optical lithography.  Future designs will reduce sensitivity to small dimensional changes by distributing capacitance over larger areas.  

To confirm that the island geometry supports photon coupling in a relevant band, we simulated the bowtie antenna response in CST Studio with a port at the junction; the resulting resonances were $\sim300 - 500$\,GHz.  Future work will focus on optimizing and measuring the SQUAT antenna sensitivity.  

\begin{table*}[t]
    \centering
    \caption{Expected design values from HFSS simulations are listed for the three SQUATs multiplexed on each chip.  The SQUATs all have nominal fin length 110\,$\mu$m, opening angle $7\pi/8$, and separation to the RF feedline of 60\,$\mu$m.  The qubit frequencies were tuned by adjusting the spacing between islands and adjusting the length of the fins, as shown in the table. The distance to the charge lines also varied between qubits, with the A, B, and C qubits having charge line spacings of 248, 254, and 260\,${\mu}$m.} \label{tab:designValues}
    \begin{tabular}{|c|c|c|c|c|c|c|c|c|c|c|}
    \hline
        SQUAT Device & $l_{gap}$ ($\mu$m) & $l_{fin}$ ($\mu$m) & $f_0$ (GHz) & $\chi$ (MHz) & $E_J/h$ (GHz) & $L_J$ (nH) & $E_C/h$ (MHz) & $C_{\Sigma}$ (fF) & $Q_c$ & $T_1$ (ns) \\
        \hline
         A & 16 & 110 & 9.31 & 13.27 & 16.3 & 10 & 780 & 35.0 & 1200 & 130 \\
         B & 13 & 111.5 & 9.11 & 9.77 & 16.3 & 10 & 745 & 36.7 & 1300 & 140 \\
         C & 10 & 113 & 8.84 & 6.47 & 16.3 & 10 & 698 & 39.1 & 1400 & 160 \\
         \hline
    \end{tabular}
\end{table*}

\section{Device Fabrication}\label{app:fab}

We fabricated the SQUATs demonstrated here on a 2" diameter, 430 $\mu$m thick annealed C-plane sapphire wafer.  The wafer was annealed for 2 hours at a maximum temp of 1200\textdegree C. The initial choice of sapphire was motivated both by minimizing charge noise for first prototypes to help make characterization easier, as compared to Si, and for consistency with the existing qubit fabrication process for other devices made by Schuster lab.

First, we deposited a 100\,nm niobium film for the ground plane. The deposition was done with a Plassys MEB550S evaporator at a pressure of 2.5e-8\,Torr, using a Ti getter before every deposition. Ground plane features were patterned with optical photolithography. Photoresist SPR3612 was manually spun on a Headway Coater and a Heidelberg MLA 150 Direct Writer was used to pattern the design.  The resist was developed in MF26A, and features were subsequently dry etched with a Plasma Therm Versaline LL ICP Metal Etcher using SF6 and CF4.  Photoresist was stripped with Remover PG at 80\textdegree C.

Next, we deposited a second optical layer to form the capacitor fins.   Photoresist SPR3612 was spun and patterned in the Heidelberg.  Photoresist was developed in MF27A.  A layer of 200\,nm thick aluminum was deposited using the Plassys evaporator at a pressure of 2.3e-8\,Torr, and liftoff was completed in Remover PG at 80\textdegree C.

For devices discussed in this paper, we fabricated the Al/AlOx/Al junctions using a Dolan process. For earlier fabrication iterations, we also investigated a 90\textdegree~Manhattan process. The Dolan-style process was chosen for two main reasons, both motivated by improving the efficiency of quasiparticle tunneling across the junction. First, the Dolan process allows fabrication of relatively wide junctions, and we found in Ref.~\cite{fink2024} that tunneling efficiency correlates strongly with junction width. Second, Dolan-style junctions provide a shorter path for quasiparticles to diffuse between the fins and junction, improving the tunneling efficiency for quasiparticles originating in the fins.

Dolan junctions were patterned with a Raith EBPG 5200+ Electron Beam Lithography System on a bi-layer of MMA EL 13 and PMMA 950 A4.  Electra was spun for charge dissipation.  The junctions were evaporated in the Plassys evaporator.  The first layer is composed of 45\,nm aluminum deposited at -23\textdegree.  The chamber was filled with an O$_2$/Ar mix and static oxidation was run at 30\ Torr.  A second layer of aluminum was deposited with 115nm thickness at angle +23\textdegree.  The wafer was diced into 1\,cm square chips with a DISCO Wafer Saw and lifted off in 80\textdegree C Remover PG.  

Test junction resistances were probed with a Micromanipulator6000 IV-CV prove station.  Finally, the chip was mounted and wirebonded with a West Bond 7476E Wedge Bonder.

\section{Experimental Setup}\label{app:exp_setup}

The results in this paper are based on measurements of three SQUAT chips, each with three qubits coupled to a single transmission line.  The first two chips, used for most measurements presented here, were operated in a Bluefors XLD400 dilution refrigerator at the SLAC Millikelvin Facility. The remaining chip, tested in parallel in a nearby Oxford ProteoxMX, was used as a third data point to verify the reliability of the device fabrication; it exhibited consistent trends in resonance frequency, dispersion, and quality factor. One of the nine qubits on the three chips did not appear, likely due to ESD damage during room temperature probing.

\subsection{Device Environment}\label{app:device_env}

Within the cryostat, details of the device environment significantly affected background parity-switching rate. Each SQUAT chip was mounted in a standard qubit-style housing (see e.g. Ref.~\cite{li2023errorcorrection}). The qubit housing was then placed inside a high-permeability enclosure to further reduce ambient magnetic fields and light leakage. The initial enclosure used for SQUATs consisted of an oxygen-free high thermal conductivity (OFHC) copper can surrounded by a tightly fit Cryoperm can. An OFHC copper lid with SMA bulkhead feedthroughs covered the top of the can. The seam between can and lid was covered using copper tape. With this shielding, the parity rate was too high to measure. In subsequent tests, we used a black-absorber lined Cryoperm can with light-tight lid designed to mitigate black-body leakage. The absorber is a mixture of carbon black, silica powder, and silicon-carbide grains suspended in Stycast epoxy~\cite{barends_cryoperm_can_coating}, supplied by collaborators in the Plourde group.

\begin{figure}
    \centering
    \includegraphics[width=0.7\linewidth]{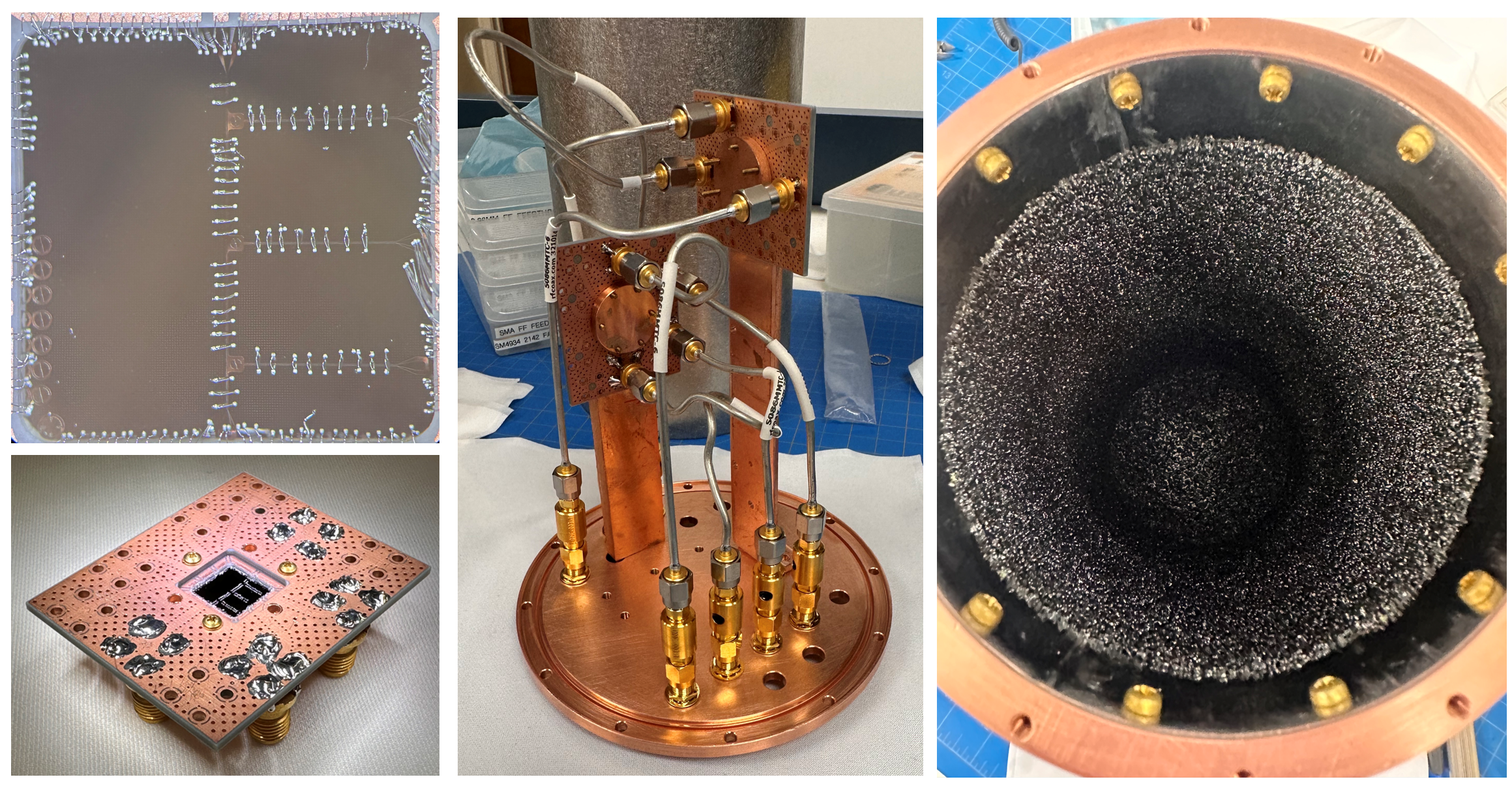}
    \caption{\textbf{Top left:} top-down view of the SQUAT chip wire bonded into the housing, shown zoomed out in bottom left. The transmission line runs from top to bottom, with the three charge lines on the right hand side. Wire bonds are used to bridge the ground plane over central conductors of feed lines. The ground plane itself is wire bonded to the ground plane of the housing around the device perimeter. A set of tests junctions can be seen in the bottom left of the chip. \textbf{Bottom Left:} device housing showing two of the three charge lines bonded to RF ports, with the transmission line connections on either end. \textbf{Middle:} Two SQUATs packaged and mounted to the underside of the IR shielding can lid. A copper cover with an indium seal covers the face of the chip. Eccosorb filters on the inside can lid are seen at the bottom of the photograph. On the lid, a cutout for a stub filter and lip to block line of sight photons can be seen. The diameter of this lip is matched to the inner can diameter, and some amount of force is required to mate them together. \textbf{Right:} Top-down view of IR-black coated Cryoperm can showing the can mated to the flange using interior brass screw to block direct line of sight. Tapped screw holes to mount to the lid are seen on the flange.}
    \label{fig:devicePackage}
\end{figure}

Only the can lid, made of OFHC copper for thermal conductivity, had low permeability. To reduce IR loading from higher temperature stages, the lid was designed without line-of-sight paths into the can. The lid also implements a free-space stub-filter to reduce photon transmission along the waveguide formed by the sealing interface~\cite{yyStubFilter}. Copper tape was again used to cover all seams. The new enclosure (shown in Fig.~\ref{fig:devicePackage}) provided a substantial reduction in parity-switching rates, as discussed in Sec.~\ref{sec:parity_meas}.

\begin{figure}
    \centering
    \includegraphics[width=0.8\linewidth]{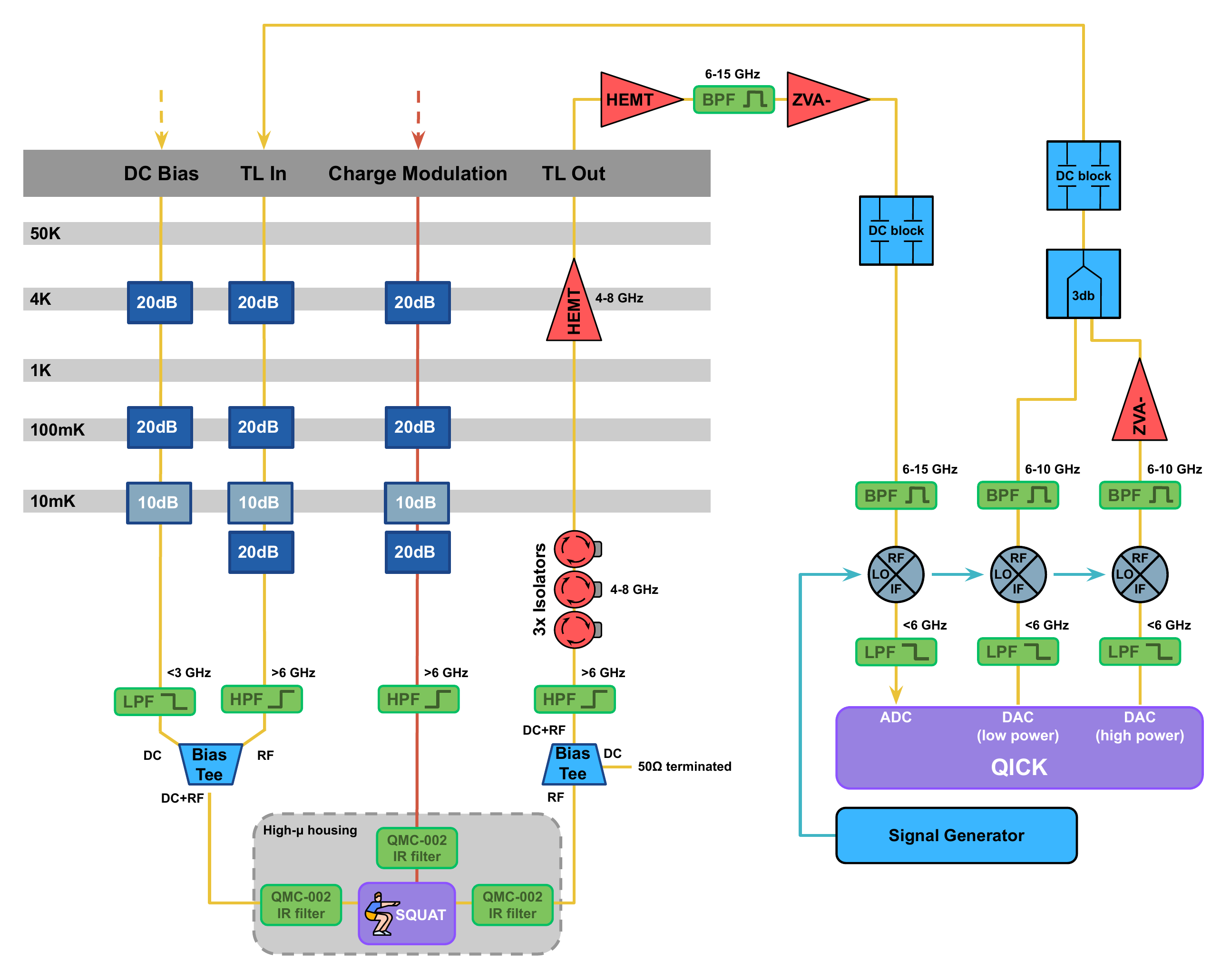}
    \caption{A typical measurement setup used to characterize SQUATs in the Bluefors XLD400. Pulsed and CW signals were generated using one of multiple QICK DACs prepared with different amplification. The choice of DAC depending on the power requirements of the measurement. RF mixers were used to reach the $\sim$8\,GHz SQUAT frequencies. Inside the fridge, bias tees were used to safely voltage bias the transmission line, providing a simple method to charge bias multiple SQUATs simultaneously. Loaded-dielectric IR filters were placed on all device ports to reduce background rate from thermal radiation. Output isolation was used to redirect and dissipate upstream amplifier noise. Device response was down-mixed and recorded using QICK ADCs. Replacing everything beyond the DC blocks with a Rohde \& Schwarz ZNB26 gives the VNA measurement setup.}
    \label{fig:readout_chain}
\end{figure}

Within the high-permeability can, each SQUAT port was shielded from thermal radiation traveling along RF coaxes using magnetically loaded dielectric-absorber Eccosorb filters, as shown in Fig.~\ref{fig:readout_chain}. Our standard filter for test applications has been the commercially available QMC-CRYOIRF in lengths of 0.6--2\,cm \footnote{\url{https://quantummicrowave.com/wp-content/uploads/2023/01/eccosorb_filter_application_note.pdf}}, which allows us to maintain low loss on the output line up to 10\,GHz. It has been shown that such filters most effectively reduce parity-switching rate when placed immediately adjacent to a device (see e.g. Ref.~\cite{serniak18, Nho_2025}). Therefore, although they are magnetic, we keep these filters inside the shielding cans alongside the SQUAT chips. Additional non-Eccosorb RF filtering is used outside the can to further reduce out-of-band noise along RF lines.

On the input lines, we used 70\,dB attenuation (plus cable loss) to achieve appropriate power on chip while reducing noise from readout electronics. On each output line, three stages of isolation were used to dissipate noise traveling back from cryogenic high electron mobility transistor (HEMT) amplifiers. We used LNF-LNC4-8 cryogenic HEMT amplifiers, with a quoted noise temperature of 1.8\,K in the 4-8\,GHz band.

Cryogenic bias tees were used to apply DC voltages to the SQUAT without damaging the RF readout electronics. In early tests, the tees were connected to charge lines, but were subsequently moved to the input line to allow for simultaneous biasing of all SQUATs through the transmission line. This was the primary method of charge biasing used for the studies in this paper. Charge bias was supplied with a Yokogawa GS200 Voltage Source.

\subsection{Measurement Setup}\label{app:meas_setup}

Most CW spectroscopy was performed using a Rohde \& Schwarz ZNB26 vector network analyzer (VNA). For parity-rate streaming, this VNA allowed for studies of fidelity vs integration time (inverse $f_{bw}$), but had limited application for continuous high-frequency sampling, as the readout mode was limited to a fixed number of samples independent of measurement length. In future studies, we plan to use the SLAC Microresonator RF (SMuRF) system \cite{Yu_2023}, which shows comparable streaming performance to the VNA and will allow for continuous multi-tone streaming.

Pulsed measurements, including those performed to characterize SQUAT $T_1$ and $T_2^*$, were taken with a Xilinx RF-SoC ZCU216 control board using the Quantum Instrumentation Control Kit (QICK, \cite{Stefanazzi_2022}) package. With QICK, the RF-SoC is capable of sending calibrated pulse strings and reading out multiple ADC inputs with excellent timing resolution and synchronization. The inclusion of multiple DACs and ADCs allows flexible readout arrangements, which was particularly useful when handling the differing power requirements of various measurements. QICK is also open source and simplifies the development of scripts to send pulses, record responses, and average with minimal timing jitter.

\section{SQUAT Readout Model}\label{app:hamiltonian} 

To understand SQUAT readout dynamics, we model the device as a two-level system coupled to an open waveguide. Our treatment follows the formalism presented in Chapter~5 of Ref.~\cite{emely_2021} with additional generalizations to include pure dephasing noise and radiative loss into non-feedline channels.  This model allows us to analytically map the qubit state to measurable transmission parameters. This model is similar to the approach taken in Refs.~\cite{sultanov,Peropadre_2013}, but we make the loss terms in this model explicit.

\subsection{Effective Hamiltonian}

We begin with a Hamiltonian that describes the qubit, the transmission line (treated as a continuum of bosonic modes), and a coupling term allowing for interaction between them:
\begin{equation}\label{eq:hamiltonian}
    H_\text{system} = H_\text{qubit} + H_\text{TL} + H_\text{int}
\end{equation}
where (using standard $\sigma$ notation for the Pauli operators)

\begin{itemize}
    \item $H_\text{qubit} = \frac{\omega_0}{2} \sigma_z$ is the Hamiltonian of a two-level system with transition frequency $\omega_0$.
    
    \item $H_\text{TL} = \int_{-\infty}^\infty d\omega J \, \omega b^\dagger_{\omega} b_{\omega}$ is the transmission line, represented as a continuum of harmonic oscillator modes with number operator $b^\dagger b$, weighted by an energy-independent density of states ($J$) with units of inverse angular frequency.
    
    \item $H_\text{int} = \int_{-\infty}^\infty d\omega J \, g \left( b^\dagger_{\omega} \sigma_- + b_{\omega} \sigma_+ \right)$ describes the coupling between the qubit and the transmission line with an energy-independent coupling ($g$) with units of angular frequency.
\end{itemize}

Using Heisenberg equations of motion and integrating over the continuum of harmonic oscillator modes, we obtain standard input-output relations:
\begin{equation}\label{eq:inout}
    \alpha_{out}^{L,R}(t) = \alpha_{in}^{R,L} (t) - i \frac{\Gamma_c}{2g} \sigma_- 
\end{equation}
where $\Gamma_c$ is the rate of qubit decay to the transmission line, defined as $\Gamma_c = 2\pi J g^2$. Here $\alpha_{out}$ and $\alpha_{in}$ are the complex amplitudes of the drive tone injected at one of the input ports, and the indices $L$ and $R$ define the transmission line port. For simplicity, we can assume our signal comes only from the left port (port 1). If we measure transmission to the right port (port 2) we find 
\begin{equation}
    S_{21} = \frac{\langle \alpha_{out}^R\rangle}{\langle \alpha_{in}^{L}\rangle}.
\end{equation}

Similarly, if we measure reflection of our signal back into the left port (port 1), we find 
\begin{equation}
    S_{11} = \frac{\langle \alpha_{out}^L \rangle}{\langle \alpha_{in}^L \rangle} = S_{21}-1.
\end{equation}

Eq.~\ref{eq:inout} makes explicit the relationship between the qubit dynamics and the reflected/transmitted field amplitudes. Since the qubit-TL interaction is mediated by the $\sigma_-$ operator, we note that our direct readout of the SQUAT is performed in the XY plane of the Bloch sphere.  This differs from typical, resonator-coupled dispersive transmon readout, which projects the qubit state onto the Z-axis (finding the expectation value $\langle  \sigma_z \rangle$). Implications and intuition for this readout mode are discussed in App.~\ref{app:pulsed_ro_diagrams}.

In the rotating frame of a coherent drive tone at frequency $\omega_d$, the qubit Hamiltonian can be written as:
\begin{equation}\label{coh_drive_H_sig+-}
    H_\text{eff} = \frac{\Delta}{2} \sigma_z + \frac{\Omega}{2} \sigma_x
\end{equation}
where (in this appendix only) $\Delta = \omega_d - \omega_0$ is the detuning between the drive and qubit, and $\Omega=2 g |\alpha^L_\text{in}|$ is the qubit Rabi frequency.

To incorporate dissipation and dephasing processes, we model qubit state evolution using the Linblad master equation.  The evolution of the qubit density matrix $\rho(t)$ is governed by:
\begin{equation}\label{eq:lindblad}
    \frac{d}{dt} \rho(t) = -i \left[ H_{\mathrm{eff}} , \rho \right] + \Gamma_{r} \mathcal{D} [\sigma_-] \rho + \frac{\Gamma_\phi}{2} \mathcal{D} [\sigma_z] \rho
\end{equation}
where $\mathcal{D}[A]$ is the Linbladian superoperator (see Sec.~5.2 of Ref.~\cite{emely_2021} for more details). Eq.~\ref{eq:lindblad} describes qubit evolution in the presence of the following decay pathways:
\begin{itemize}
    \item $\Gamma_r$ is the radiative decay rate, encompassing all $T_1$-style decoherence processes.  We can further break this down into feedline (`coupled') and non-feedline (`loss') decay paths $\Gamma_r = \Gamma_c + \Gamma_l$
    \item $\Gamma_\phi$ is the pure dephasing rate
\end{itemize}
From this master equation, one can derive the time-evolution equations for $\sigma_{\pm}$ and $\sigma_z$:
\begin{align}
    \dot{\langle \sigma_{\pm} \rangle} &= \left(\pm i\Delta - \gamma\right)\langle \sigma_{\pm} \rangle \mp i\frac{\Omega}{2}\langle \sigma_z \rangle \label{eq:phaseEvolution} \\
    \dot{\langle \sigma_z \rangle} &= -i\Omega \left(\langle \sigma_{+} \rangle - \langle \sigma_{-} \rangle\right) - \Gamma_r\left(\langle \sigma_z \rangle + 1 \right) \label{eq:stateEvolution} 
\end{align}
where the total decay rate $\gamma$ is given by
\begin{equation}
    \gamma \equiv \frac{1}{2}\Gamma_{r} + \Gamma_\phi = \frac{1}{2} (\Gamma_c + \Gamma_l ) + \Gamma_\phi
\end{equation}
These time-evolution equations can be solved in two limits. In the transient limit, we can find the decoherence ($T_1$) and dephasing ($T_2$) times, and make measurements as one would a conventional dispersively coupled qubit, with modifications to the measurement scheme discussed in Sec.~\ref{sec:pulsed_meas}. In the continuous limit, the time-evolution can be set to zero, and we find steady-state expectation values as a function of drive frequency and strength.

\subsection{Transient Response}

We find the transient response first for phase decay, setting $\Omega=0$ in Eq.~\ref{eq:phaseEvolution} to get
\begin{equation}
    \dot{\langle \sigma_{\pm} \rangle} = \left(\pm i\Delta - \gamma\right)\langle \sigma_{\pm} \rangle.
\end{equation}
This has the solution
\begin{equation}
    \langle \sigma_{\pm} \rangle (t) = C\exp{\left[\left(\pm i\Delta - \gamma\right)t\right]} = e^{-\gamma t}\left[c_1\cos(\Delta t) \pm c_2\sin(\Delta t)\right]
\end{equation}
Thus we have two counter-rotating states with precession frequency equal to the detuning, even in the limit that $\gamma$ vanishes. If we define $T_2$ to be the time during which the envelope of this function reaches 1/e, we thus find
\begin{equation}
    \gamma T_2 = 1 \rightarrow T_2 = \frac{1}{\gamma}
\end{equation}
This tells us that the total decay rate we've defined is the inverse $T_2$ time of the SQUAT when measured using the transient response.

Now we can find $T_1$ in a similar manner using Eq.~\ref{eq:stateEvolution}. Setting $\Omega$ to zero, we find
\begin{equation}
    \dot{\langle \sigma_z \rangle} = - \Gamma_r\left(\langle \sigma_z \rangle + 1 \right)
\end{equation}
We take the time derivative of this equation to get the ODE
\begin{equation}
    \frac{d^2}{dt^2} \langle {\sigma_z} \rangle + \Gamma_r \frac{d}{dt}\langle \sigma_z \rangle = \frac{d}{dt} (\frac{d}{dt}+\Gamma_r)\langle \sigma_z \rangle) = 0
\end{equation}
This tells us there are two solutions: one where the time derivative is zero, which we recognize occurs from the initial equation when $\langle \sigma_z \rangle = -1$ (the qubit ground state). The nontrivial solution is 
\begin{equation}
    \langle \sigma_z \rangle = 2P_1(t=0)e^{-\Gamma_r t} - 1
\end{equation}
where $P_1(t=0)$ is the probability of initially finding the system in the excited state. Following the same approach as before, we find
\begin{equation}
    \Gamma_r T_1 = 1 \rightarrow T_1 = \frac{1}{\Gamma_r}
\end{equation}
This confirms that $T_1$ is the inverse radiative decay rate.

Finally, we use these definitions to estimate $\Gamma_{\phi}$

\begin{equation}
    \Gamma_{\phi} = \gamma - \frac{1}{2}\Gamma_r = \frac{1}{T_2} - \frac{1}{2 T_1} 
\end{equation}

In other words, if radiative loss dominates system dynamics ($\gamma\approx \Gamma_r/2$) we recover $T_2\approx2T_1$.

\subsection{Steady-State Response}

We can also solve for the steady-state expectation values of the excitation operators, which is needed to solve Eq.~\ref{eq:inout}. Setting $\dot{\langle \sigma_{\pm} \rangle} = \dot{\langle \sigma_z \rangle} = 0$ in Eqs.~\ref{eq:phaseEvolution} and \ref{eq:stateEvolution} yields the solution
\begin{equation}
    \langle \sigma_- \rangle = - \frac{\Gamma_r}{2} \frac{\Omega(\Delta+i\gamma)}{ \Omega^2 \gamma +\Gamma_r\Delta^2+\Gamma_r\gamma^2}
\end{equation}
which yields the transmission and reflection equations, using Eq.~\ref{eq:inout},
\begin{align} \label{HM_S21}
        S_{21} &= 1 - \frac{\Gamma_c}{2\gamma} \frac{1-i\frac{\Delta}{\gamma}}{1 + \left(\frac{\Delta}{\gamma} \right)^2 + \frac{\Omega^2}{\gamma \Gamma_r}}\\
        S_{11} &= - \frac{\Gamma_c}{2\gamma} \frac{1-i\frac{\Delta}{\gamma}}{1 + \left(\frac{\Delta}{\gamma} \right)^2 + \frac{\Omega^2}{\gamma \Gamma_r}}
\end{align}
This is the same result derived in Refs.~\cite{sultanov,Peropadre_2013}, with the caveat that instead of $\Gamma_c$ in the denominator of the drive term, we find $\Gamma_r$. This allows for the distinction between total radiative loss and loss specifically to the transmission line, which is significant in the case that we drive the SQUAT through the charge line, which has a smaller coupling.

Following the example of Ref.~\cite{Peropadre_2013}, we can equivalently express the Rabi rate in terms of drive photon interrogation rate~($\Gamma_n$) or the tone's on-chip power~($P_r$): 
\begin{equation}\label{eq:PeropadreEquation2}
    \frac{\Omega^2}{\Gamma_r} = 2\Gamma_n = \frac{4 P_r}{hf}
\end{equation}
This allows us to relate the resonance equation to physical measurement parameters.
\begin{align} \label{HM_S21_alt}
        S_{21} &= 1 - \frac{\Gamma_c}{2\gamma} \frac{1-i\frac{\Delta}{\gamma}}{1 + \left(\frac{\Delta}{\gamma} \right)^2 + \frac{2\Gamma_n}{\gamma}}\\
        S_{11} &= - \frac{\Gamma_c}{2\gamma} \frac{1-i\frac{\Delta}{\gamma}}{1 + \left(\frac{\Delta}{\gamma} \right)^2 + \frac{2\Gamma_n}{\gamma}}
\end{align}

In the low-power limit ($2\Gamma_n/\gamma \ll 1$), the SQUAT response reduces to
\begin{equation}
S_{21} =  1 - \frac{\Gamma_c}{2\gamma}\frac{1}{1+i\frac{\Delta}{\gamma} }
\end{equation}
We can relate this to the equation used to fit superconducting resonators, which takes the following form (see Ref.~\cite{GaoThesis}):
\begin{equation}
    S_{21} = 1 -\frac{\frac{Q_r}{Q_c}e^{i\phi}}{1+2iQ_r\Delta/\omega_r}
\end{equation}
This suggests the following mapping:
\begin{equation}\label{Quality_factors}
    Q_c = \frac{\omega_0}{\Gamma_c} = \frac{\omega_0}{\Gamma_r - \Gamma_l}, \;\; Q_r = \frac{\omega_0}{2\gamma}= \frac{\omega_0}{2}T_2
\end{equation}
For a well-designed SQUAT with small internal losses, we find that $\Gamma_r \approx \Gamma_c$ and $Q_c \approx \omega_0 T_1$.  We can also compute $Q_i$:
\begin{equation}
    Q_i = \left(Q_r^{-1} - Q_c^{-1}\right)^{-1} = \frac{\omega_0}{\Gamma_l + 2\Gamma_{\phi}}
\end{equation}
Therefore, in the low-power limit, we may use standard resonator characterization tools to quantify all system parameters.  By fitting the resonance as a function of readout power, we can also verify that the inferred on-chip power is linear with respect to the applied power (see Fig.~\ref{fig:fscan_all}).

\subsection{Qubit Excitation}\label{app:excitation}

We can also use Eqs.~\ref{eq:phaseEvolution} and \ref{eq:stateEvolution} with $\dot{\langle \sigma_{\pm} \rangle} = \dot{\langle \sigma_z \rangle} = 0$ to solve for the steady-state value of $\langle \sigma_z \rangle$:

\begin{equation}
    \langle \sigma_z \rangle = -1 +  \frac{\frac{2\Gamma_n}{\gamma}}{1 + \left(\frac{\Delta}{\gamma} \right)^2 + \frac{2\Gamma_n}{\gamma}} 
\end{equation}

This allows us to calculate the expected excitation probability ($P_1$) of a driven qubit.

\begin{align}
    P_1 &= \frac{1+\langle \sigma_z \rangle}{2}\\
    &= \frac{1}{2} \frac{\frac{2\Gamma_n}{\gamma}}{1 + \left(\frac{\Delta}{\gamma} \right)^2 + \frac{2\Gamma_n}{\gamma}} 
\end{align}

The result is consistent with expectations. When $\Gamma_n \rightarrow0$, there is no readout-induced excitation. When $\Gamma_n \rightarrow\infty$, the excitation probability reaches its maximum of 1/2.

\section{Pulsed Readout Interpretation}\label{app:pulsed_ro_diagrams}

The SQUAT is directly coupled to a transmission line rather than dispersively coupled to a readout resonator.  As a result, measurement does not project the qubit state onto the energy eigenbasis (Z~basis), but instead probes coherences in the equatorial (XY) plane of the Bloch sphere.  This section provides physical intuition for this effect and presents Bloch-sphere schematics for the pulsed measurement sequences described in the main text.

In a continuous measurement, coherent microwave photons propagate along the transmission line and interact with the qubit. A fraction of these photons are absorbed, exciting the qubit, and are subsequently re-emitted back into the transmission line.  These re-emitted photons acquire a phase shift relative to photons that propagate past the qubit without interaction.  The measured signal consists of the coherent superposition of these two fields: photons that interacted with the qubit, and photons that did not.  The resonance features observed in Fig.~\ref{fig:fscan_all} arise from the interference between these two contributions.  Importantly, the measured transmission does not directly count excitations lost to the environment; instead, it is sensitive to the phase and amplitude of the coherently re-emitted field.  As explained in App.~\ref{app:hamiltonian}, this field is proportional to the expectation value of the qubit lowering operator $\langle \sigma_-\rangle$, which depends on the qubit's phase coherence rather than its population.  Consequently, states aligned with the Z~axis ($\ket{0}$ and $\ket{1}$) emit no coherent radiation, while states in the XY~plane produce maximal signal.

In this framework, the readout effectively converts phase information of the qubit into amplitude variations of the transmitted field.  Manipulations that rotate the qubit within the XY~plane therefore modify the interference condition and directly modulate the measured transmission amplitude.  The measurement can be interpreted as a projection onto the equatorial plane of the Bloch sphere, rather than a projective energy measurement.

\subsection{Dephasing Measurement ($T_2^*$)}

\begin{figure}[ht]
    \centering
    \includegraphics[height=2.5cm]{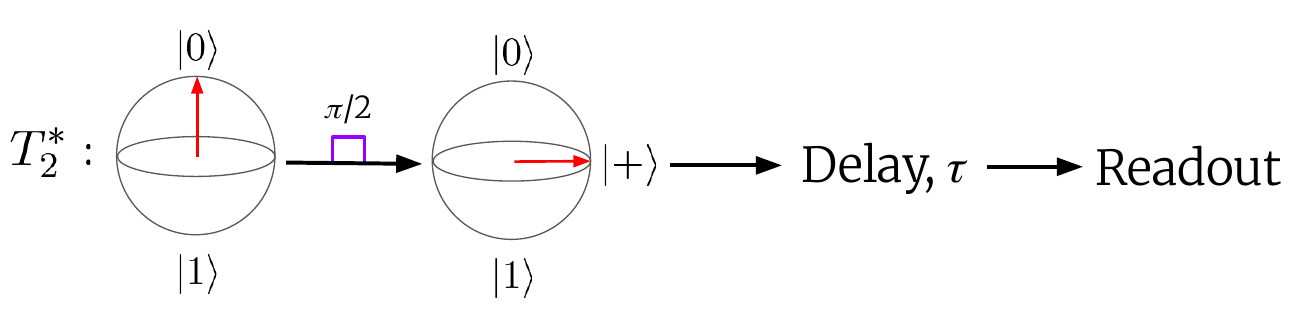}
    \caption{The $T_2^*$ measurement sequence. The qubit is initialized in the ground state and rotated into the equatorial plane with a $\pi/2$ pulse.  During a variable free-evolution delay ($\tau$), transverse coherence decays and the state accumulates phase at a rate set by the detuning ($\Delta$) between the drive and qubit frequencies.  The remaining coherence is measured via the emitted transmission-line field within a short integration window.}
    \label{fig:pulse_sequence_T2}
\end{figure}

The dephasing measurement sequence is drawn in Fig.~\ref{fig:pulse_sequence_T2}.  The qubit is initialized in the ground state, then a calibrated $\pi/2$ pulse is used to rotate the qubit into $\ket{+}$.  Next, during a subsequent variable delay time~($\tau$), the Bloch vector precesses about the Z~axis at a rate set by the detuning~($\Delta$) between the drive and qubit frequencies.  On resonance, this precession is suppressed, and the transverse coherence decays with the characteristic time $T_2^*$.

\begin{figure}[ht]
    \centering
    \includegraphics[width=0.8\linewidth]{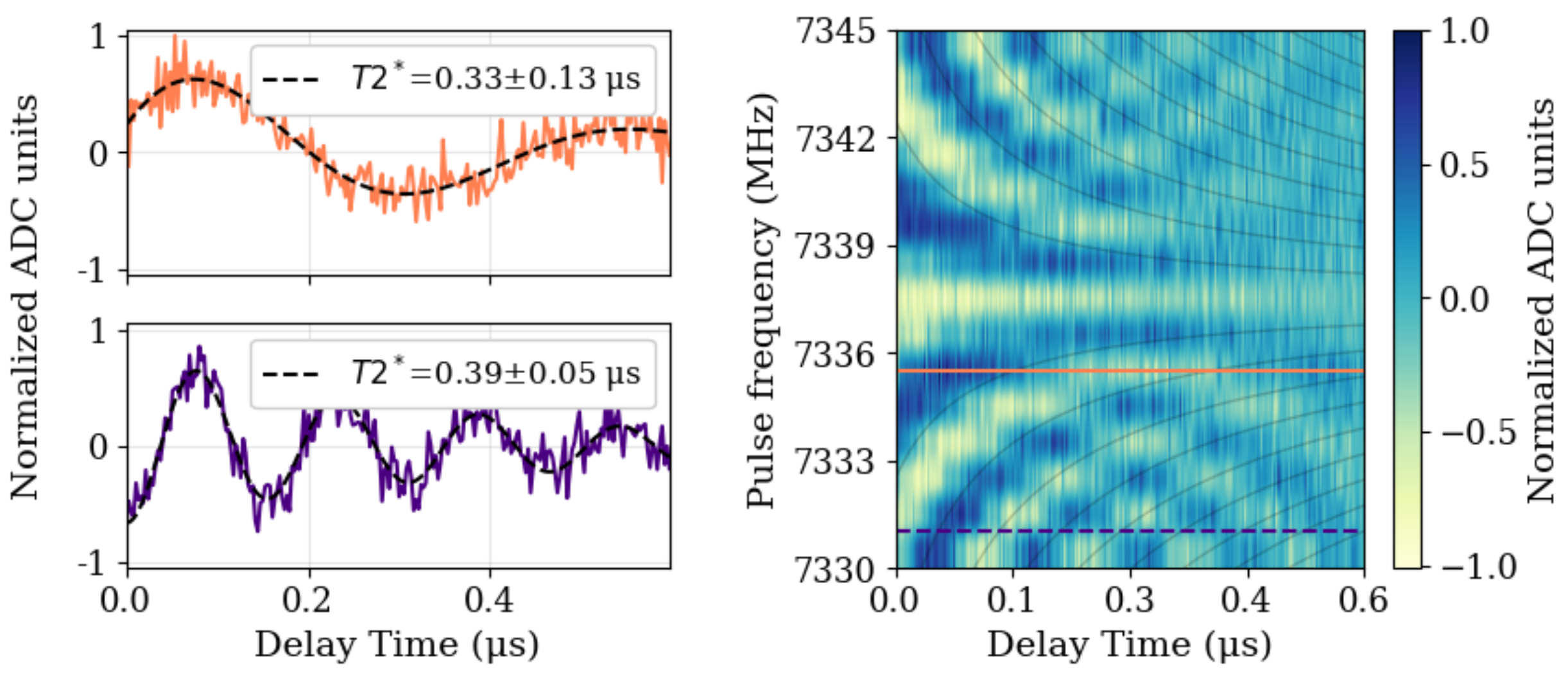}\\
    \caption{
    \textbf{Left Top:} A representative time-domain trace taken small drive/qubit detuning.  The oscillations superimposed on the decay are at the frequency of this detuning.  Dashed curves show fits to a decaying cosine model, from which the dephasing time $T_2^*$ is extracted.  
    \textbf{Left Bottom:} A representative time-domain trace at a large drive/qubit detuning.  The higher oscillation frequency allows for reduced fit uncertainty.
    \textbf{Right:} A heatmap of the qubit emission to the feedline, plotted as a function of both free-evolution delay time and drive frequency.  Oscillations arise from coherent precession of the Bloch vector in the equatorial plane due to detuning between the drive and qubit frequencies.  Grey lines indicate contours of constant accumulated phase, corresponding to fixed phase offsets (increments of $\pi/2$) between the drive and qubit during the delay interval.  These contours visualize the expected timing of maxima and minima in the transverse signal as a function of detuning.  Colored horizontal lines mark the drive frequencies for the selected time-domain traces shown on the left.  The orange (solid) line corresponds to the top left subplot, and the purple (dashed) line corresponds to the lower left subplot.}
    \label{fig:T2_freq}
\end{figure}

Because the readout measures $\sigma_-$, the decay of the transverse coherence appears directly as a decay in the emitted field amplitude.  By recording the probability of detecting a photon within a fixed, short time window following the delay, and repeating this measurement as a function of $\tau$, we can plot the dephasing time.  When the qubit is detuned from the drive tone, however, oscillations at the frequency $\Delta$ are convolved with the trace.  When the drive is slightly detuned (or if the parity bands are not perfectly degenerate), the period of these oscillations is often comparable to the coherence time.  This can lead to accidental underestimation of $T_2^*$ during measurement.  To mitigate this effect, one may prefer to run the measurement at an intentionally larger $\Delta$, such that the oscillations are fully visible in the trace and can be removed via fitting.  An example of $T_2^*$ measurements as a function of drive frequency is given in Fig.~\ref{fig:T2_freq}.

\subsection{Energy Relaxation Measurement ($T_1$)}

\begin{figure}[ht]
    \centering
    \includegraphics[height=2.5cm]{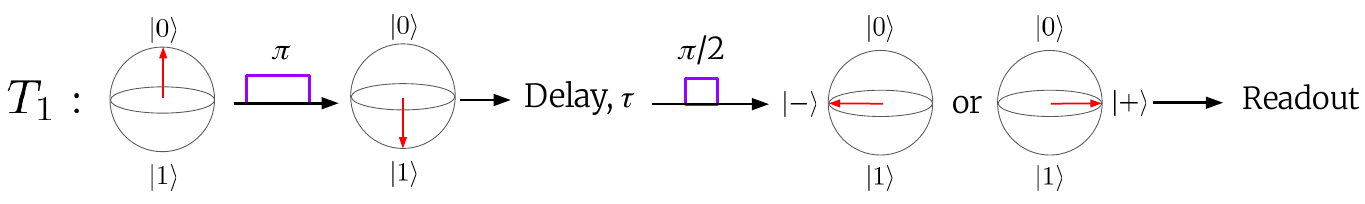}
    \caption{The $T_1$ measurement sequence. The qubit is first prepared in the excited state using a $\pi$ pulse.  This is followed by a variable delay time ($\tau$).  A subsequent $\pi/2$ pulse maps the remaining excited-state population onto transverse coherence, which is measured via the emitted transmission-line signal within a short integration window.}
    \label{fig:pulse_sequence_T1}
\end{figure}

Measuring energy relaxation in the SQUAT configuration requires an additional rotation to map population differences onto transverse coherence.  The qubit is first excited to $\ket{1}$ using a $\pi$ pulse and allowed to evolve freely for a variable delay time ($\tau$), during which relaxation may occur. A subsequent $\pi/2$ pulse rotates the Bloch vector into the equatorial plane, converting the remaining excited-state population into a measurable transverse component.  Following this rotation, the emitted field is again measured within a short integration window. If the qubit has relaxed to $\ket{0}$ during the delay, the final rotation produces a state with opposite phase compared to the case where the qubit remained excited. Averaging over repeated trials therefore yields a signal proportional to the excited-state population, which decays exponentially with time constant $T_1$.  This measurement sequence is drawn in Fig.~\ref{fig:pulse_sequence_T1}.  As in the dephasing measurement, detuning introduces oscillations due to residual XY~plane precession.

Based on the frequency-dependent fits described in Sec.~\ref{sec:deviceOnly}, the SQUAT devices studied here exhibit decay times on the order of 100\,ns, significantly shorter than those of conventional resonator-coupled transmons.  With the use of $\sim$10s\,ns control pulses, the measurement is sensitive to timing jitter at the nanosecond scale.  Additionally, at elevated quasiparticle parity-switching rates, the qubit transition frequency fluctuates on timescales comparable to a single measurement.  This introduces an effective random detuning between the drive and qubit during the pulse sequence, leading to additional dephasing and reduced contrast in the measured oscillations. These effects limit the achievable fidelity of pulsed measurements and are consistent with the observed decay rates reported.

\section{SQUAT Readout Fidelity} \label{app:fidelity}

Readout fidelity ($\mathcal{F}$) for a SQUAT was derived in Ref.~\cite{fink2024} assuming response could be described as a conventional resonator modified by the efficiency of interrogating the SQUAT, which becomes vanishingly small in the high-power limit. In this appendix, we confirm the approximation holds in the low-power limit, and find modified behavior at higher readout powers.  We do so using the two-level-system model of $S_{21}$ from Eq.~\ref{eq:S21Squat} and fully define $\mathcal{F}$ in Sec.~\ref{app:fidelity_def}, below.

We consider two potential readout scenarios, amplitude readout and phase readout. Here we describe both separately, as they require different optimizations for the same device. In both cases, we compute SNR in terms of the input/output voltages ($\tilde{V}_{in}$ and $\tilde{V}_{out}$) starting with the relation
\begin{equation}
    P_{in,out} = \frac{|\Tilde{V}_{in,out}|^2}{Z_0} \rightarrow |\tilde{V}_{in,out}| = \sqrt{P_{in,out}Z_0}
\end{equation}
with $Z_0$ the characteristic impedance of the transmission lines used in our measurements, and $P_{in, out}$ are input/output powers. This allows us to model the noise contribution as thermal voltage noise and to describe transmission as the voltage received at the output.

First we consider the noise term. For a noise temperature $T_n$, the voltage noise (with no signal present) has the variance 
\begin{equation}
    \sigma_V^2 = 4 k_b T_n f_{bw}\eta\left(\beta^{-1}\right)Z_{0}
\end{equation}
where $\eta$ is the quantum correction 
\begin{equation}
    \eta\left(x\right) = \frac{x}{\exp(x)-1}+\frac{x}{2}
\end{equation}
with $\beta=\frac{k_BT_n}{hf}$.  The voltage noise variance reduces to the standard quantum limit in the case that $k_b T_n\rightarrow 0$. In the limit $k_bT_n \gg hf$, $\eta$ approaches unity.

Now we consider the transmitted output voltage.
\begin{equation}
    \Tilde{V}_{out}  = \Tilde{S}_{21}\Tilde{V}_{in}
\end{equation}
Our signal is the difference between the even and odd states, so we have
\begin{equation}
    \Delta \Tilde{V} = \Tilde{V}_{out,even} - \Tilde{V}_{out,odd} = \left(\Tilde{S}_{21,even}-\Tilde{S}_{21,odd}\right)\Tilde{V}_{in}
\end{equation}
The noise term has no phase information, so to compute SNR (the power signal-to-noise ratio) we need to find the magnitude squared of this difference vector, which becomes
\begin{equation}
    \Delta V^2 = |\Delta \Tilde{V}|^2 =  \left|\left(\Tilde{S}_{21,even}-\Tilde{S}_{21,odd}\right)\right|^2|\Tilde{V}_{in}|^2 = \left|\left(\Delta\Tilde{S}_{21}\right)\right|^2 P_{in}Z_0
\end{equation}
The normalized noise in the space where the even and odd states are separated by a unit vector is thus
\begin{equation}
    \sigma^2_s \equiv \mathrm{SNR}^{-1} = \frac{2\sigma_V^2}{\Delta V^2} = 2\left|\left(\Delta\Tilde{S}_{21}\right)\right|^{-2}\frac{4 k_b T_n f_{bw}\eta\left(\beta^{-1}\right)}{P_{in}}
\end{equation}
If we change variables to write in terms of the interrogation rate ($\Gamma_n = 2P_{in}/hf$), we find the final equation:
\begin{equation}
    \sigma^2_s = 16\left|\left(\Delta\Tilde{S}_{21}\right)\right|^{-2}\beta\frac{f_{bw}}{\Gamma_n}\eta\left(\beta^{-1}\right)
\end{equation}

\subsection{Amplitude (Dissipative) Readout}

First, let's consider amplitude readout. In this case, we read out on one parity state ($\delta f=0$), and assume the other is sufficiently displaced that we can approximate transmission in that state to be 1 ($2\chi \gg \gamma$). We then use Eq.~\ref{HM_S21_alt} to describe $\Delta S_{21}$ in terms of $\Gamma_n$. In this case, we find: 
\begin{equation}
    \Delta |S_{21}| \approx 1 - \left|1 - \frac{\Gamma_c}{2\gamma} \frac{1}{1 + 2 \frac{\Gamma_n}{\gamma}}\right| = \frac{\Gamma_c}{2\gamma} \frac{1}{1 + 2 \frac{\Gamma_n}{\gamma}}
\end{equation}
In the low-power limit ($2\Gamma_n/\gamma \ll 1$), $\Delta S_{21}$ is equivalent to the diameter of the resonance circle, as predicted when using the conventional resonance equation. Without making that assumption, the state variance becomes:
\begin{equation}\label{eq:intermediate-sigma}
    \sigma^2_s = 16\beta \frac{(2\gamma)^2}{\Gamma_c^2}\frac{(1+2 \frac{\Gamma_n}{\gamma})^2}{\Gamma_n}f_{bw}\eta\left(\beta^{-1}\right)
\end{equation}
This equation is useful to optimize the amplitude readout with respect to drive power. We find that variance is minimized when $\Gamma_n$ is
\begin{equation}\label{eq:pwr_opt_phase_ro}
  \bar{\Gamma}_n = \frac{\gamma}{2} = \frac{1}{2T_2}
\end{equation}
In an ideal readout, the measurement bandwidth is equivalent to the inverse integration window.  We can express this interval as a number of qubit $T_2$ times.  We substitute
\begin{align}\label{eq:N-def}
    f_{bw}=(N2T_2)^{-1} = (2N/\gamma)^{-1}
\end{align}
into Eq.~\ref{eq:intermediate-sigma} to find a power-optimized state variance of
\begin{equation}
    \bar{\sigma}^2_s = 64\frac{\beta}{N} \frac{(2\gamma)^2}{\Gamma_c^2}\eta\left(\beta^{-1}\right).
\end{equation}
Finally, in the ideal limit of zero dephasing and zero non-feedline loss ($\Gamma_r = \Gamma_c = 2\gamma$), we recover $T_2=2T_1$ and find a best-case state variance of
\begin{equation}\label{eq:AmplitudeOptimum}
    \bar{\sigma}^2_s = 64\frac{\beta}{N}\eta\left(\beta^{-1}\right)
\end{equation}
which is identical to the statement that there is a white-noise dominated resolution, which decreases as $\frac{1}{N}$ due to averaging. 

The above is analogous to the derivation in Ref.~\cite{fink2024}, with the additional benefit of providing the optimal readout power for amplitude readout of an ideal device.
\begin{equation}
    \bar{P}_{in} \approx hf\frac{ \bar{\Gamma}_n}{2} = hf\frac{\gamma}{4} \approx hf\frac{\Gamma_r}{8} = \frac{hf}{8 T_1}
\end{equation}
This is exactly the `magic power' referred to in Ref.~\cite{amin2024}, which is equivalent to driving the SQUAT at roughly one photon per decay time in the case of $\Gamma_{\phi}\approx 0$.

\subsection{Phase (Dispersive) Readout}

We now consider readout with frequency $f$ centered between the parity states ($\delta f_{even} = -\delta f_{odd} = \chi$, where $2\chi$ is the frequency difference between the even and odd states). In this mode, the magnitude of the transmitted signal is unchanged by a parity operation, and information lives entirely in the signal phase. In this case, we find:

\begin{equation}
    \Delta |S_{21}| \approx \left|  \frac{\Gamma_c}{2\gamma} \frac{1-i\frac{\chi}{\gamma}}{1 + \left(\frac{\chi}{\gamma} \right)^2 + 2 \frac{\Gamma_n}{\gamma}}   -   \frac{\Gamma_c}{2\gamma} \frac{1+i\frac{\chi}{\gamma}}{1 + \left(\frac{\chi}{\gamma} \right)^2 + 2 \frac{\Gamma_n}{\gamma}}   \right| = \frac{\Gamma_c}{2\gamma} \frac{2\frac{\chi}{\gamma}}{1 + \left(\frac{\chi}{\gamma} \right)^2 + 2 \frac{\Gamma_n}{\gamma}}
\end{equation}
which gives us the normalized state variance
\begin{equation} \label{eq:phase_SNR}
    \sigma^2_s = 16\beta \frac{(2\gamma)^2}{\Gamma_c^2}\frac{(1+\chi^2/\gamma^2+2\Gamma_n/\gamma)^2}{(2\chi /\gamma)^2 \Gamma_n}f_{bw}\eta\left(\beta^{-1}\right)
\end{equation}
We can jointly optimize along both $\Gamma_n$ and voltage-tunable $\chi$.  We start with $\Gamma_n$:
\begin{equation}
    \bar{\Gamma}_n \equiv \frac{\gamma}{2} \left( 1+\chi^2/\gamma^2 \right)
\end{equation}
which approaches the optimum found for amplitude readout when $\chi \ll \gamma$.  Substituting the full expression for $\bar{\Gamma}_n$ back into Eq.~\ref{eq:phase_SNR}, we find:
\begin{equation}
    \sigma^2_s = 16\beta \frac{(2\gamma)^2}{\Gamma_c^2}\frac{2(1+\chi^2/\gamma^2)}{\chi^2/\gamma}f_{bw}\eta\left(\beta^{-1}\right)
\end{equation}
which shows the surprising behavior that SNR is maximized by driving the dispersion to its highest value, while jointly raising readout power.  In the limit of sufficiently high dispersion ($\chi \gg \gamma$), we achieve the optimal SNR
\begin{equation}
    \bar{\sigma}^2_s = 16\beta \frac{(2\gamma)^2}{\Gamma_c^2}\frac{2}{\gamma}f_{bw}\eta\left(\beta^{-1}\right)
\end{equation}
We again substitute $f_{bw}=(N2T_2)^{-1} = (2N/\gamma)^{-1}$ to find the power-optimized state variance of
\begin{equation}
    \bar{\sigma}^2_s = 16\frac{\beta}{N} \frac{(2\gamma)^2}{\Gamma_c^2}\eta\left(\beta^{-1}\right)
\end{equation}
Finally, we take the limit of zero dephasing and zero non-feedline loss ($\Gamma_r = \Gamma_c = 2\gamma$), giving
\begin{equation}\label{eq:PhaseOptimum}
    \bar{\sigma}^2_s = 16\frac{\beta}{N}\eta\left(\beta^{-1}\right).
\end{equation}
This is a factor of 4 in variance and 2 in voltage improvement over the amplitude readout case, and suggests a potential way forward for designing SQUATs with higher power-handling capabilities.  However, since we cannot operate at arbitrarily high readout powers, it is worth considering an optimization with respect to our other tunable variable ($\chi$) for a finite value of $\Gamma_n$. Starting again from Eq.~\ref{eq:phase_SNR}, we find
\begin{equation}\label{eq:chi_opt_phase_ro}
    \bar{\chi} = \gamma \sqrt{1 + \frac{2\Gamma_n}{\gamma}}
\end{equation}
Giving a dispersion-optimized variance of
\begin{equation}
    \sigma^2_s = 16\beta \frac{(2\gamma)^2}{\Gamma_c^2}\frac{1+2 \Gamma_n/\gamma}{\Gamma_n}f_{bw}\eta\left(\beta^{-1}\right)
\end{equation}
If we optimize for power, we again find infinite power gives the same global minimum variance in Eq.~\ref{eq:PhaseOptimum}. However, the improvement becomes marginal above our previously defined high-power limit ($2\Gamma_n/\gamma \gg 1$).

\begin{figure}
    \centering
    \includegraphics[width=0.9\linewidth]{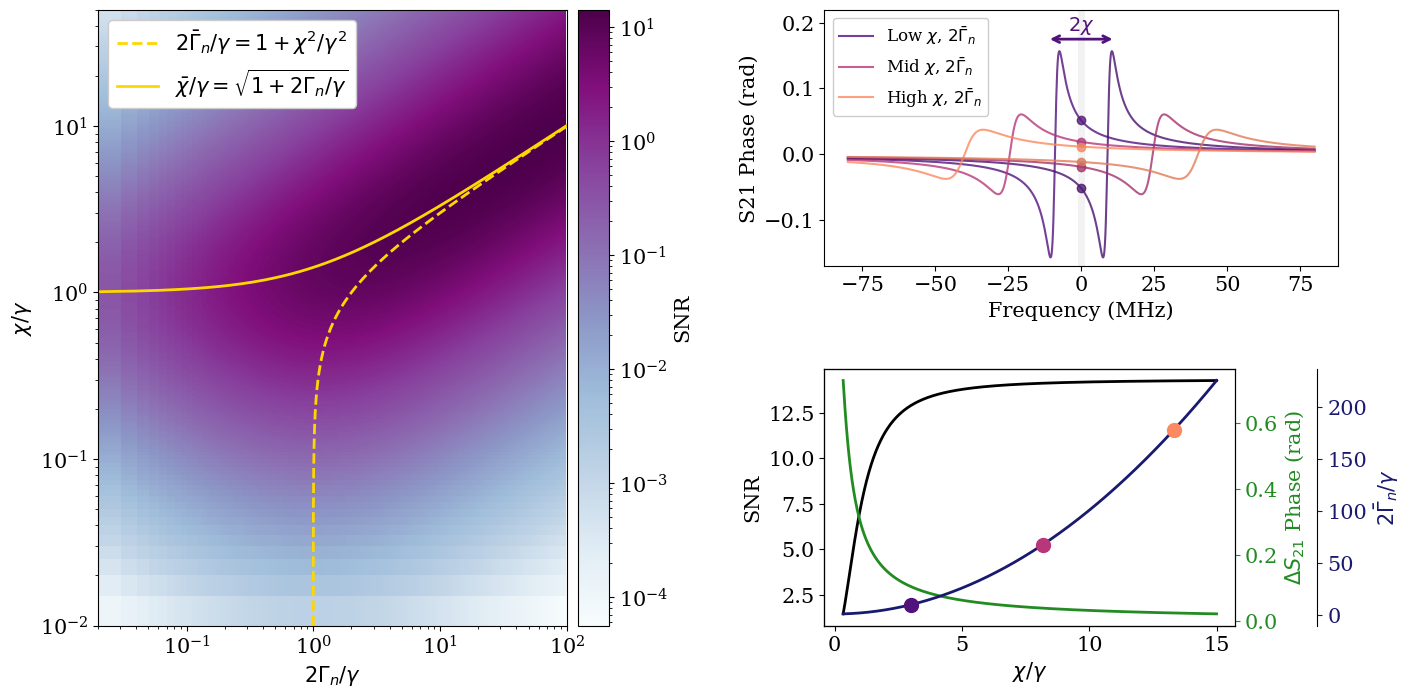}
    \caption{Plots to visualize the expected relationship between readout power, dispersion, variance, and SNR for phase readout. The following parameter choices were used:  $T_\text{eff}=2$\,K, $\gamma=3$\,MHz, $\Gamma_c=2\gamma$, $f_0=8$\,GHz, and $f_{bw}=5$\,kHz.
    \textbf{Left:} Phase readout SNR, plotted as a function of $\chi/\gamma$ and $2\Gamma_n/\gamma$.   For any $\chi$, the dashed yellow curve gives the optimal value of $\Gamma_n$ ($\bar{\Gamma}_n$).  For any $\Gamma_n$, the solid yellow curve gives the optimal value of $\chi$ ($\bar{\chi}$).  
    \textbf{Top Right:} 
    Phase of the complex transmission ($S_{21}$) plotted as a function of frequency for three representative values of $\chi$, evaluated at the optimized $\bar{\Gamma}_n$ (as given by Eq.~\ref{eq:pwr_opt_phase_ro}).  For each $\chi$, the parity bands are plotted separately.  A vertical line marks $f_0$, the bare qubit frequency and the location of the readout tone.  Points mark the transmission phase at $f_0$ for each parity state.  The difference between the two points gives the phase of $\Delta S_{21}$, the change in signal resulting from a parity switch.  At low $\chi$, lower readout power is preferred, giving a large $\Delta S_{21}$ (but higher $\sigma_s$ due to the relative contribution of noise in low-power readout). $\bar{\Gamma}_n$ increases with $\chi$, leading to smaller $\Delta S_{21}$, but relatively lower noise from higher readout power.
    \textbf{Bottom Right:} As a function of $\chi$, we plot SNR, $\Delta S_{21}$, and $\bar{\Gamma}_n$.  The three representative $\chi$ from the top right subplot are shown as points on the $2\bar{\Gamma}_n/\gamma$ trace.  We see SNR increase with increasing $\chi$ (due to the quadratic increase in $\bar{\Gamma}_n$) but plateau at higher $\chi$ (due to the inverse relationship between $\Delta S_{21}$ and $\chi$).}
    \label{fig:SNR_plot}
\end{figure}

We plot phase-readout state variance as a function of power and dispersion in Fig.~\ref{fig:SNR_plot}, along with optimization lines for $\Gamma_n$ with constant $\chi$, and $\chi$ with constant $\Gamma_n$.  The two optimization lines converge around $\Gamma_n/\gamma \sim 10$ and $\chi/\gamma\sim 3$.  

\subsection{Readout Fidelity Definition}\label{app:fidelity_def}

We can now relate state variance due to readout noise with readout fidelity ($\mathcal{F}$).  This is relevant to our analysis of parity-switching rates, which are fit from PSDs of the time-domain data using Eq.~\ref{eq:fidelity_main} (copied here for ease of readability).
\begin{equation}\label{eq:parity}
    \text{PSD}(f) = \mathcal{F}^2\frac{4\Gamma_{p}}{(2\Gamma_{p})^2+(2\pi f)^2} + (1-\mathcal{F}^2)f_{bw}^{-1}
\end{equation}
$\Gamma_{p}$ is the parity-switching rate and $f_{bw}$ is the readout bandwidth described in the previous section.  In Ref.~\cite{Rist__2013}, $\mathcal{F}$ is defined relative to the probability of improper state reconstruction ($p_\mathrm{false}$). We use this definition for consistency with our parity-switching analysis.

\begin{equation}
    p_\mathrm{false} = \frac{1}{2}(1-\mathcal{F})
\end{equation}

This is reasonable, since perfect fidelity ($\mathcal{F}\rightarrow1$) causes $p_\mathrm{false} \rightarrow 0$, and zero fidelity ($\mathcal{F}\rightarrow0$) gives an equal chance of reconstructing the even or odd state ($p_\mathrm{false} \rightarrow 1/2$).  We can equate the reconstruction-error probability due solely to the noise contribution as the probability that a random sample from a Gaussian with variance $\sigma_s^2$ exceeds 0.5 (half the distance between the states). This is given by the equation
\begin{equation}
    p_\mathrm{false} = \int_{1/2}^{\infty}\frac{dx}{\sqrt{2\pi\sigma_s^2}}e^{\frac{-x^2}{2\sigma_s^2}} = \frac{1}{2}\left[1-\mathrm{erf}\left(\frac{1}{\sqrt{8\sigma_s^2}}\right)\right]
\end{equation}
which shows that
\begin{equation}\label{eq:sig2F}
    \mathcal{F} = \mathrm{erf}\left(\frac{1}{\sqrt{8\sigma_s^2}}\right)
\end{equation}
in the limit that only noise contributes to the misidentification probability. As this is the intended regime for SQUAT operation, Eq.~\ref{eq:sig2F} provides a guideline for computing fidelity as a function of design parameters. The probability of misidentifying the state due to parity-switching events is validated in the next section using a simple Monte Carlo simulation.

We can now compute $\mathcal{F}$ using the state variance $\sigma^2_s$ from Eq.~\ref{eq:phase_SNR},
\begin{equation}
    \mathcal{F} = \mathrm{erf}\left(\frac{1}{\sqrt{8\sigma_s^2}}\right) \approx  \mathrm{erf}\left(\sqrt{\frac{\Gamma_c^2}{128\beta(2\gamma)^2}\frac{(2\chi /\gamma)^2 \Gamma_n}{f_{bw}(1+\chi^2/\gamma^2+2\Gamma_n/\gamma)^2}\eta^{-1}\left(\beta^{-1}\right)}\right).
\end{equation}
We see that fidelity for a fixed noise temperature can vary substantially with readout conditions, drive strength, and readout bandwidth. 

In the limit $k_bT_n \gg hf$, where $\eta$ approaches unity, one can show that this equation reduces to the form
\begin{align}\label{eq:F-noise}
    \mathcal{F} = \mathrm{erf}\left(\sqrt{\frac{N}{8\bar{\sigma}^2_s (T_{\text{eff}})}}\right),
\end{align}
where we have absorbed the effect of all sub-optimal readout and dispersion parameters into an effective noise temperature ($T_{\text{eff}}$). $N$ represents the number of independent averaging time periods, defined in the previous section.

\section{Parity-Switching Trends: Fidelity and Environmental Effects}\label{app:parity}

Here we provide further studies into parity-switching behavior, including the effect of background switching on measured fidelity and initial studies of switching rate drivers.

\subsection{Parity-Switching Reduction in Fidelity}

An important caveat to the simple fidelity definition in the prior appendix is that, in the $f_{bw}\sim\Gamma_{p}$ regime, additional state confusion arises from parity-switching occurring within the sample integration time. This causes the fidelity determined by PSD fits to be lower than the noise-only fidelity given in Eq.~\ref{eq:F-noise} (especially for low $f_{bw}$).  It further implies that there is an $f_{bw}$ which produces optimal $\mathcal{F}$ for a given $\Gamma_{p}$. This implication can be understood intuitively by noticing if $f_{bw}\rightarrow\infty$, then $\sigma^2_s\rightarrow\infty$ and $\mathcal{F}\rightarrow0$. Conversely, if $f_{bw}\ll\Gamma_{p}$, there are several switching events per integration time and $\mathcal{F}\rightarrow0$. Between these regimes lies the optimal $\mathcal{F}$.

\begin{figure}
    \centering
    \includegraphics[width=0.47\linewidth]{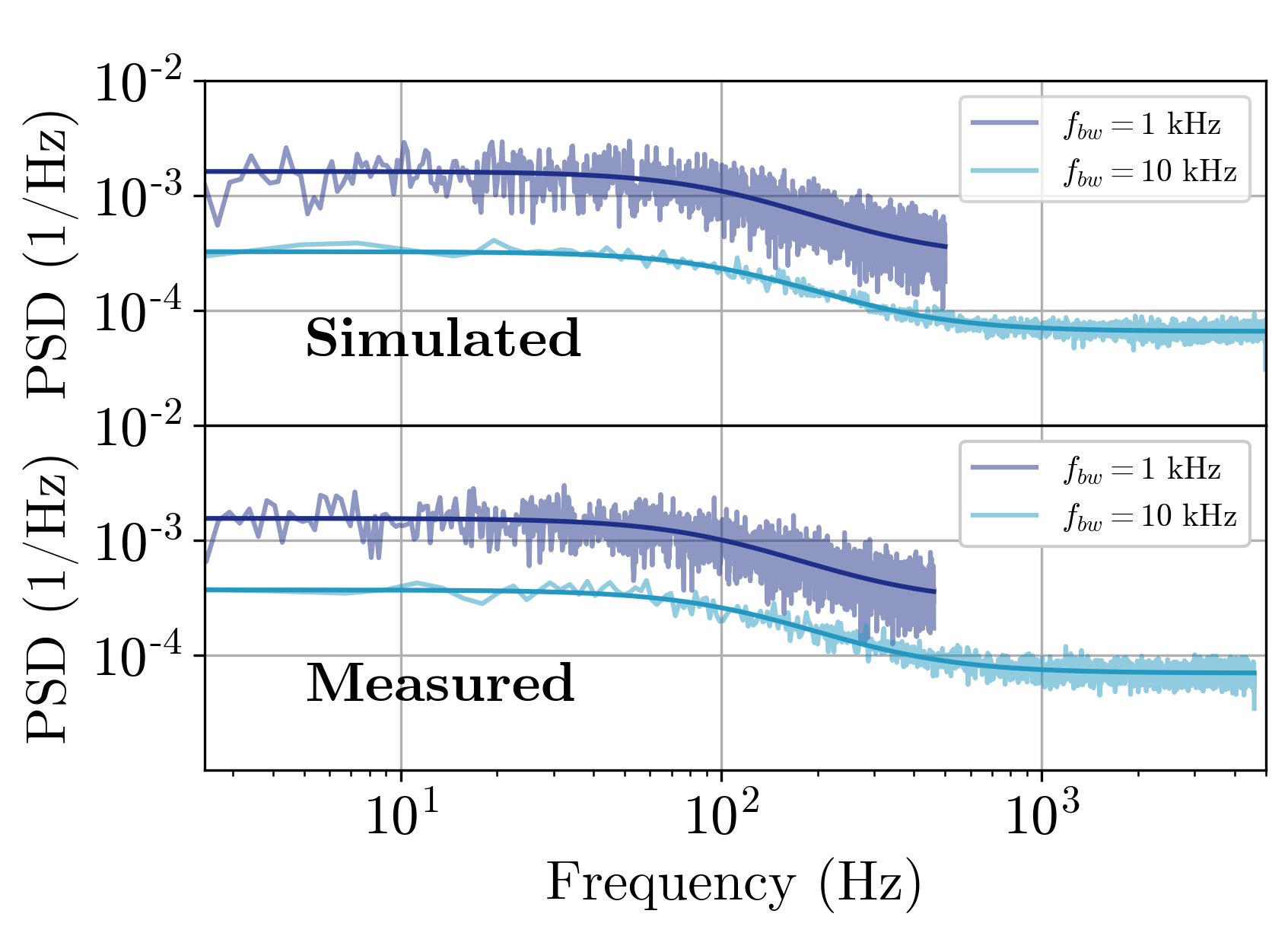}
    \includegraphics[width=0.5\linewidth]{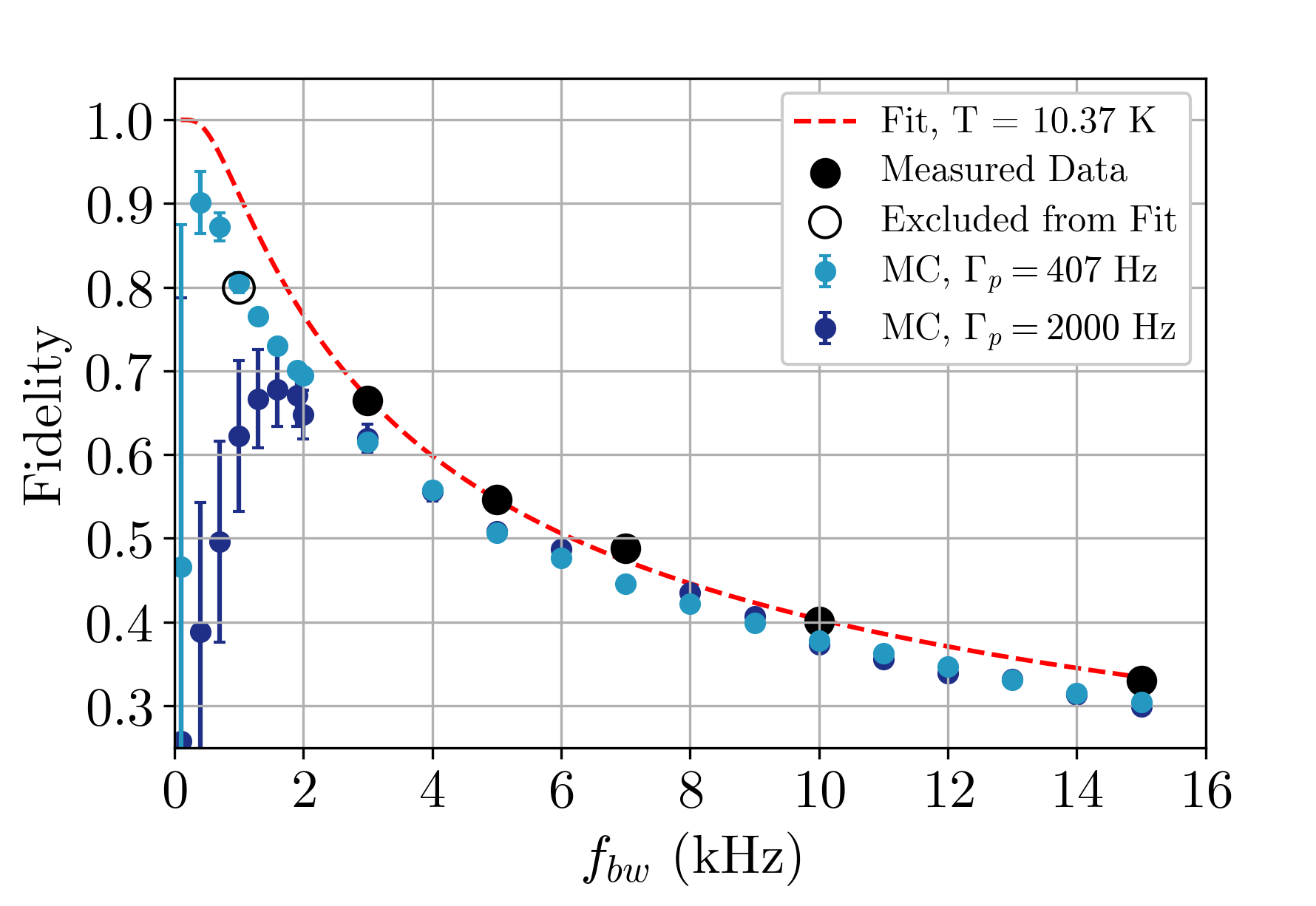}
    \caption{\textbf{Left:} Monte Carlo data and a fit to Eq.~\ref{eq:parity} to extract parity-switching rate and fidelity. For these simulations, the parity-switching rate is fixed, and the sampling rate is increased. \textbf{Right:} Comparison of resulting fidelity from a sweep of readout bandwidth compared to the analytic model (dashed) and a set of points acquired with different $f_{bw}$. The data point near 1\,kHz is excluded from the fit due to $f_{bw}$ approaching the measured switching rate ($\sim450$\,Hz). All other data agree well with a noise temperature of 10.37\,K .}
    \label{fig:MCModel}
\end{figure}

To understand this behavior, empirical SQUAT $\mathcal{F}$ was measured by fitting PSDs at several $f_{bw}$ with Eq.~\ref{eq:parity}. The resulting trends were then fit to Eq.~\ref{eq:F-noise}. When supplied with $N$ (see Eq.~\ref{eq:N-def}), the fit yields $\beta=k_bT_{\text{eff}}/hf_0$, from which an upper bound on the SQUAT noise temperature of $T_{\text{eff}}=10.37$\,K was derived. This assumes thermal or quantum limited noise is the only limitation on fidelity. The data and fit appear in the right panel of Fig.~\ref{fig:MCModel}. As discussed above, Eq.~\ref{eq:F-noise} overestimates $\mathcal{F}$ when $f_{bw}\lesssim\Gamma_{p}$. For this reason, the lowest $f_{bw}$ point is excluded from the fit.

The $\mathcal{F}$ behavior observed in measurements at low $f_{bw}$ can be understood through Monte Carlo modeling. To simulate parity-switching time-domain data of arbitrary length and frequency, the following algorithm was employed:
\begin{enumerate}
    \item The generated dataset has duration $T_\mathrm{exp}$ and time resolution $\Delta t_\mathrm{base}=f_s^{-1}$, with $f_{s}$ as the maximum sampling rate. Additionally, the output sample rate $\Delta t_\mathrm{sample}=f_{bw}^{-1}$ is specified and tuned to study behavior as a function of $f_{bw}$.
    \item The dataset is populated with samples from a Gaussian distribution with zero mean and standard deviation $\sigma_s$ computed for the sampling rate $f_{bw}=f_s$. 
    \item A sequence of switching times is produced, determined by a Poisson process with rate $\lambda=\Gamma_{p}\Delta t_\mathrm{base}$. At these switching times, the Gaussian-populated dataset is shifted by $\pm1/2$ (with alternating sign). This creates two populations with variance $\sigma_s^2$ and unitary mean separation.
    \item To simulate data acquisition at a lower sample rate (or longer duration $\Delta t_\mathrm{sample}$ between samples) than the resolution of the generated dataset ($\Delta t_\mathrm{base}$), every $N=\Delta t_\mathrm{sample}/\Delta t_\mathrm{base}$ samples are averaged together.
    \item This dataset now resembles a dataset produced via measurement of a SQUAT with $f_{bw}=1/\Delta t_\mathrm{sample}$. The PSD is calculated and $\mathcal{F}$ and $\Gamma_{p}$ are extracted via fit to equation~\ref{eq:parity}.
\end{enumerate}
The values of $\mathcal{F}$ extracted from measured data as well as from several simulated datasets of varying $f_{bw}$ and two different $\Gamma_{p}$ are displayed in the right panel of Fig.~\ref{fig:MCModel}. The light blue points are derived from datasets with underlying $\Gamma_{p}$ matching that of measured data, while $\Gamma_{p}$ of the dark blue is set to 2\,kHz to clearly demonstrate that $\mathcal{F}$ trends downwards when $f_{bw}\sim\Gamma_{p}$. A comparison of PSDs and their fits between simulated and measured data can be seen in the left panel for two different values of $f_{bw}$.

One immediate feature of note is that the Monte Carlo, when simulated according to the noise temperature extracted from data, under-predicts the fidelity of the data by a small amount. We explored a number of alternative definitions of fidelity, with others based on probability distribution products over-estimating fidelity by a small amount. This was true independent of the use of specific variance models. We thus expect a slight mismatch between our analytic fidelity and that extracted by the fits. In the absence of a more accurate derivation, this fidelity definition is close enough to that extracted from Monte Carlo that it provides a useful guide for device optimization. More work is needed to understand the discrepancies between Monte Carlo and analytic expectation.

\subsection{Environmental Sources of Parity-Switching}\label{app:vibration}

\begin{figure*}[t]
    \centering
    \includegraphics[width=0.9\linewidth]{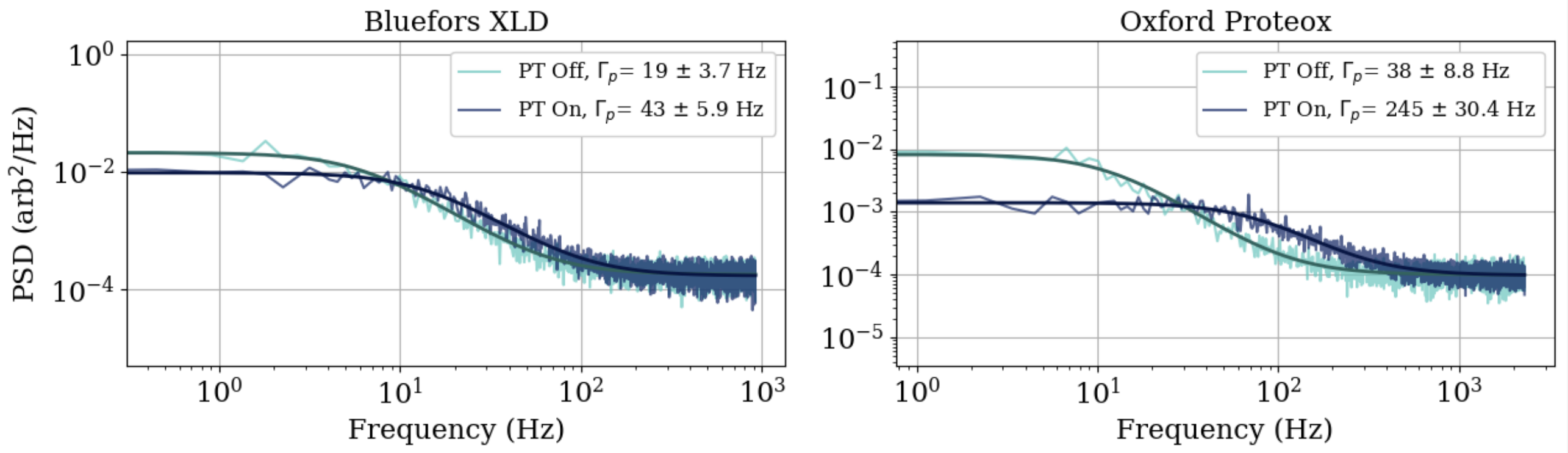}
    \caption{Power spectral densities (PSDs) of SQUAT time-domain data measured in two different cryostats.  Each plot shows data acquired with the pulse-tube on and off, together with the corresponding fits used to extract the parameters summarized in Tab.~\ref{tab:PTswitch}.}
    \label{fig:pt_on_off_bfg_olaf}
\end{figure*}

In the prototype SQUAT devices discussed here, parity-switching rates were found to be sensitive to environmental noise. The most prominent noise was that associated with operation of pulse-tube cryocoolers, which are necessary to continuously run the dry dilution refrigerators housing the SQUATs. In our system, pulse-tube noise did not introduce distinct spectral features in the measured power spectral densities.  Instead, it manifested as an increase in the steady-state parity-switching rate when the pulse-tube was running (see Fig.~\ref{fig:pt_on_off_bfg_olaf}).  The relative scale of this contribution varied with experimental conditions and is expected to depend on a combination of factors such as the cryostat model, device mounting scheme, and ambient noise environment (as characterized by $\Gamma_{\text{other}}$ and $x_{qp}^{ne}$).  A systematic characterization of environmental noise sources, including pulse-tube–induced effects and their mitigation, is beyond the scope of this work but will be the subject of future studies.

\subsection{Readout-Power Driven Parity-Switching}\label{app:readout_power}

\begin{figure*}[t]
    \centering
    \includegraphics[width=0.6\linewidth]{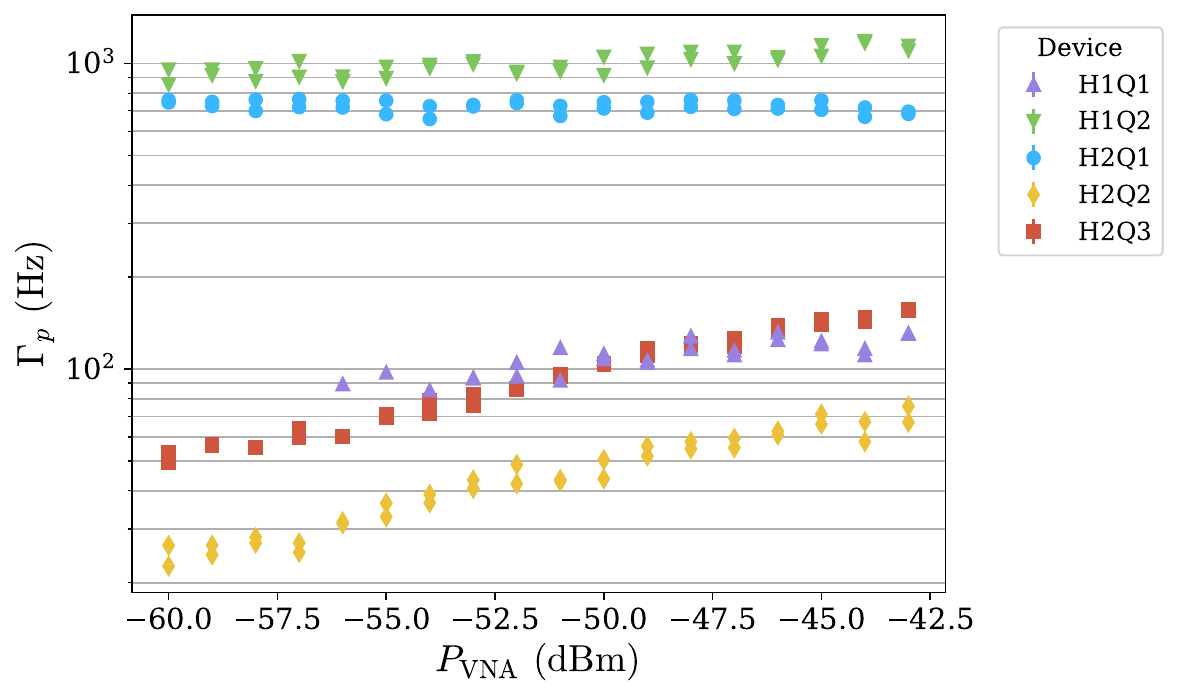}
    \caption{SQUAT parity-switching rates as a function of probe-tone readout power from the VNA ($P_\text{VNA}$). Note there is an additional 70\,dB of cryogenic attenuation between the VNA output and the devices. Rates that were low for low $P_\text{VNA}$ increased with $P_\text{VNA}$. Rates that started high remained high. }
    \label{fig:rate_vs_power}
\end{figure*}

Parity-switching rates were also observed to depend on CW readout power. Fig.~\ref{fig:rate_vs_power} shows a representative measurement.  For the lowest-switching-rate SQUATs, we observe a clear increase in switching rate with increasing readout power. This dependence is not seen in higher-rate SQUATs.  This behavior could originate from various contributing mechanisms, including increased excited-state qubit population ($P_1$) at higher drive powers and drive-induced heating in the measurement chain.  The figure is included for completeness and visualization; a detailed characterization of these underlying mechanisms is beyond the scope of the present work and will be addressed in future studies.

\end{document}